\documentclass{JHEP3}

\usepackage{amsmath,amstext,amssymb}
\usepackage{psfrag,graphicx,subfigure}
\usepackage{cite}

\def\calo{{\cal O}}

\def\calt{{\cal T}}
\def\calj{{\cal J}}
\def\ie{{i.\,e.\ }}
\def\eg{{e.\,g.\ }}

\def\mt{\tilde m}
\def\phit{\tilde \phi}

\def\calh{{\cal H}}

\def\call{{\cal L}}

\def\re{\text{Re}}
\def\im{\text{Im}}
\def\k{\kappa}

\def\del{\partial}
\def\diag{\mathrm{diag}}

\def\ee{{\mathrm e}}
\def\ii{{\mathrm i}}

\newcommand{\dd}{\mathrm{d}}

\title{Transport in Anisotropic Superfluids: A Holographic Description}

\author{Johanna Erdmenger, Patrick Kerner and Hansj\"org Zeller \\ Max-Planck-Institut f\"ur Physik (Werner-Heisenberg-Institut)\\
  F\"ohringer Ring 6, 80805 M\"unchen, Germany\\ \email{jke, pkerner, zeller@mppmu.mpg.de}}

\abstract{We study transport phenomena in p-wave superfluids in the context of gauge/gravity duality. Due to the spacetime anisotropy of  this system, the tensorial structure of the transport coefficients is non-trivial in contrast  to the isotropic case. In particular, there is an additional shear mode which leads to a non-universal value of the shear viscosity even in an Einstein gravity setup.  In this paper, we present a complete study of the helicity two and helicity one fluctuation modes. In addition to the non-universal shear viscosity, we also investigate the thermoelectric effect, i.e.~the mixing of electric and heat current. Moreover, we also find an additional  effect due to the anisotropy, the so-called flexoelectric effect.}

\date{\today}


\preprint{MPP-2011-113}

\begin{document}

\newpage


\section{Introduction}
\label{sec:Introduction}
Hydrodynamics is a very powerful description of systems close to equilibrium.  Its focus is on slowly varying fluctuations with frequency $\omega$ and momentum $k$ smaller than the typical length scale, the mean free path. Hydrodynamics  may be seen as the low-energy effective description of interacting  systems. Gauge/gravity duality is a very useful tool to further develop the hydrodynamic description for various systems. New transport phenomena have been uncovered by studying systems which violate parity by an anomaly \cite{Erdmenger:2008rm,Banerjee:2008th,Son:2009tf}.  The transport in a system which shows the chiral magnetic effect induced by an axial anomaly has been studied in \cite{Lifschytz:2009si,Gynther:2010ed,Kalaydzhyan:2011vx,Hoyos:2011us}. Effects of anisotropy in strongly coupled systems have been discussed in \cite{Mateos:2011tv,Rebhan:2011ke}. Recently the hydrodynamic description for s-wave superfluids which may violate parity has been investigated in \cite{Herzog:2011ec,Bhattacharya:2011ee,Bhattacharya:2011tr}. 

The hydrodynamical description of superfluids is interesting since an Abelian symmetry is spontaneously broken. Due to the spontaneous breaking of a continuous symmetry, a Nambu-Goldstone boson appears in the spectrum. Since it is massless, it behaves as hydrodynamic mode and has to be included into the hydrodynamical description. In this paper we study p-wave superfluids where in addition to the Abelian symmetry,  the rotational symmetry is spontaneously broken and thus more Nambu-Goldstone bosons appear in the spectrum. This leads to an anisotropic fluid in which the transport coefficients depend on the direction, \ie they are tensors.  In the case we study here the fluid is transversely symmetric, \ie the system has an $SO(2)$ symmetry and we can use this symmetry to reduce the tensors to the minimal amount of independent quantities. For instance, the viscosity which relates the stress $T^{\mu\nu}$ in a fluid with the strain $\nabla_\lambda u_\rho+\nabla_\rho u_\lambda$ given in terms of the four velocity of the fluid $u_\mu$ is parametrized by a rank four tensor $\eta^{\mu\nu\lambda\rho}$ (see \eg \cite{Landau:1959te,Gennes:1974lc}). Using the symmetry we find two independent shear viscosities, in contrast to only one in the isotropic case, \ie $SO(3)$ symmetry.

The shear mode is the transversely polarized fluctuation given for instance by $\nabla_y u_z+\nabla_z u_y$ for a momentum in $x$ direction. In the isotropic case this is the unique shear mode since any momentum  can be rotated into the $x$ direction by the $SO(3)$ rotational symmetry. In the transversely symmetric case, two momenta, one along and one perpendicular to the favored direction, \eg the $x$ direction, must be considered. Thus there are two shear modes. If the momentum is along the favored direction, the situation is similar to the isotropic case and the strain is again $\nabla_y u_z+\nabla_z u_y$. However if the momentum is perpendicular to the favored direction say in $y$ direction, the situation changes dramatically. Now the little group is given by the discrete group $\mathbb{Z}_2$ and the strain is given by $\nabla_x u_z+\nabla_z u_x$. Since the shear viscosity can be evaluated at zero momentum, we can characterize the fluctuations with respect to the full symmetry group which is in the transversely symmetric case $SO(2)$. The first fluctuation $\nabla_y u_z+\nabla_z u_y$ is a helicity two state as the shear mode in the isotropic case is. The second fluctuation $\nabla_x u_z+\nabla_z u_x$ however transform as helicity one state under the rotational symmetry. This transformation property is due to the rotational symmetry breaking and will be very important in this paper.

\medskip

In the context of gauge/gravity duality, the spontaneous breaking of continuous symmetries by black holes developing hair was first achieved in \cite{Gubser:2008px} and later used to construct holographic superconductors/superfluids by breaking an Abelian symmetry \cite{Hartnoll:2008vx,Hartnoll:2008kx}. Along this line also p-wave superconductors/superfluids have been constructed \cite{Gubser:2008wv} and gave rise to the first string theory embeddings of holographic superconductors/superfluids \cite{Ammon:2008fc,Basu:2008bh,Ammon:2009fe}. In order to obtain the effects of spontaneous rotational symmetry breaking in the hydrodynamics of p-wave superfluids, we have to take the back-reaction of the superfluid density into account, \ie we consider the effect of the superfluid density on the energy-momentum tensor. This was obtained \eg in \cite{Ammon:2009xh}. On the gravity side, the p-wave superfluid state corresponds to an asymptotically AdS black hole which carries vector hair.

A very famous result in the context of gauge/gravity duality is that the ratio between shear viscosity and the entropy density is universal \cite{Kovtun:2004de,Buchel:2003tz,Iqbal:2008by}. The ratio is the same for all field theories which have a Einstein gravity dual, \ie the field theory is a large $N$ gauge theory at infinite 't Hooft coupling $\lambda$. This universality can be proven as follows: The shear mode is the only mode which transforms as a helicity two mode under the $SO(2)$ little group and thus decouple from all the other modes. In addition it can be shown that the low energy dynamics of this mode is trivial such that the ratio is completely determined by gravitational coupling constant which is universal. The universality is lost if finite $N$ and/or coupling is considered for instance by adding a Gauss-Bonnet term to the gravity action (see \eg \cite{Buchel:2008vz,Cremonini:2011iq}). 

In the letter \cite{Erdmenger:2010xm} we have shown  that universality is also absent even at leading order in $N$ and $\lambda$ if the fluid is anisotropic. In this case, the universality is lost since one of the different shear modes transforms as a helicity one mode under the rotational symmetry and can therefore couple to other helicity one modes present in the system. The coupling generates non-trivial dynamics which lead to a non-universal behavior of the shear viscosity. This result is valid for a field theory dual to Einstein gravity without additional contributions to the gravity action. In this paper we present the detailed calculations for  this result. This calculation was suggested already in \cite{Natsuume:2010ky}. We study the complete set of the helicity two and one modes in the $SU(2)$ Einstein-Yang-Mills theory in the broken phase at zero momentum.

Along this calculation we find some additional transport phenomena: the thermoelectric effect in the transversal directions and the flexoelectric effect. The thermoelectric effect is the phenomenon that the electric and heat current mix since charged object transport charge as well as energy. This effect has already been studied for holographic s-wave superfluids \cite{Hartnoll:2008kx, Hartnoll:2009sz}. We find that the thermoelectric effect in the transversal directions agrees with the result found for s-wave superfluids. The flexoelectric effect is known from nematic liquids which consists of molecules with non-zero dipole moment (see \eg \cite{Gennes:1974lc}). A direction can be preferred by the dipoles. In this anisotropic phase, a strain can lead to effective polarization of the liquid and an electric field applied to the liquid can lead to a stress. This is the first appearance of this effect in the context of gauge/gravity duality.

The paper is organized as follows: In section~\ref{sec:Holographic-Setup-and} we review the holographic setup in which p-wave superfluids are constructed and describe their behavior in equilibrium.  In section~\ref{sec:Perturbations-about-the} we study perturbations about equilibrium. We characterize the fluctuations in terms of their transformation under the symmetry groups and determine their equations of motion. In addition we calculate the on-shell action and read off the correlation functions. In section~\ref{sec:Transport-Properties} we extract the transport properties out of the correlation functions and find the non-universal shear viscosity, the thermoelectric effect and the flexoelectric effect.  We conclude in section~\ref{sec:Conclusion}. In the appendix~\ref{sec:Holographic-Renormalization} we discuss holographic renormalization. The gauge covariant fields are constructed in appendix~\ref{sec:Constructing-the-Gauge}.  In appendix~\ref{sec:Numerical-Evalutation-of} we review the numerical evaluation of correlator when operator mixing is present.  Some general remarks on anisotropic fluids are given in appendix~\ref{sec:General-Remarks-on}.


\section{Holographic Setup and Equilibrium}
\label{sec:Holographic-Setup-and}
We consider $SU(2)$ Einstein-Yang-Mills theory in $(4+1)$-dimensional asymptotically AdS space. The action is
\begin{equation}
\label{eq:action}
S = \int\!\dd^5x\,\sqrt{-g} \, \left [ \frac{1}{2\k_5^2} \left( R -\Lambda\right) - \frac{1}{4\hat g^2} \, F^a_{MN} F^{aMN} \right] + S_{\text{bdy}}\,, 
\end{equation}
where $\k_5$ is the five-dimensional gravitational constant, $\Lambda = - \frac{12}{L^2}$ is the cosmological constant, with $L$ being the AdS radius, and $\hat g$ is the Yang-Mills coupling constant. The $SU(2)$ field strength $F^a_{MN}$ is
\begin{equation}
F^a_{MN}=\del_M A^a_N -\del_N A^a_M + \epsilon^{abc}A^b_M A^c_N \,,
\end{equation}
where capital Latin letter as indices run over $\{t,x,y,z,r\}$, with $r$ being the AdS radial coordinate, and $\epsilon^{abc}$ is the totally antisymmetric tensor with $\epsilon^{123}=+1$. The $A^a_M$ are the components of the matrix-valued gauge field, $A=A^a_M\tau^a dx^M$,  where the $\tau^a$ are the $SU(2)$ generators, which are related to the Pauli matrices by $\tau^a=\sigma^a/2\ii$.  $S_{\text{bdy}}$ includes boundary terms that do not affect the equations of motion, namely the Gibbons-Hawking boundary term as well as counterterms required for the on-shell action to be finite. Finally it is convenient to define
\begin{equation} \label{eq:alpha}
\alpha \equiv \frac{\kappa_5}{\hat g} \, ,
\end{equation}
which measures the strength of the backreaction.

The Einstein and Yang-Mills equations derived from the above action are
\begin{align}
\label{eq:einsteinEOM}
R_{M N}+\frac{4}{L^2}g_{M N}&=\k_5^2\left(T_{MN}-\frac{1}{3}{T_{P}}^{P}g_{MN}\right)\,, \\
\label{eq:YangMillsEOM}
\nabla_M F^{aMN}&=-\epsilon^{abc}A^b_M F^{cMN} \,,
\end{align}
where the Yang-Mills stress-energy tensor $T_{MN}$ is
\begin{equation}
\label{eq:energymomentumtensor}
T_{M N}=\frac{1}{\hat{g}^2}\left(F^a_{PM}{F^{aP}}_{N}-\frac{1}{4}g_{MN} F^a_{PQ}F^{aPQ}\right)\,.
\end{equation}

\subsection{Hairy Black Hole Solution}
\label{sec:Hairy-Black-Hole}
Following ref. \cite{Gubser:2008wv,Ammon:2009xh}, to construct charged black hole solutions with vector hair we choose the gauge field ansatz
\begin{equation}
\label{eq:gaugefieldansatz}
A=\phi(r)\tau^3\dd t+w(r)\tau^1\dd x\,.
\end{equation}
The motivation for this ansatz is as follows. In the field theory we will
introduce a chemical potential for the $U(1)$ symmetry generated by $\tau^3$.
We will denote this $U(1)$ as $U(1)_3$. The bulk operator dual to the $U(1)_3$
density is $A^3_t$, hence we include $A^3_t(r) \equiv \phi(r)$ in our ansatz.
We want to allow for states with a nonzero $\langle \calj^x_1\rangle$, so in
addition we introduce $A^1_x(r) \equiv w(r)$. With this ansatz for the gauge
field, the Yang-Mills stress-energy tensor in eq.~\eqref{eq:energymomentumtensor} is
diagonal. Solutions with nonzero $w(r)$
will preserve only an $SO(2)$ subgroup of the $SO(3)$ rotational symmetry, so
our metric ansatz will respect only $SO(2)$. In addition the system is invariant under the $\mathbb{Z}_2$ parity transformation $P_\|$: $x\to -x$ and $w\to -w$. Furthermore, given that the
Yang-Mills stress-energy tensor is diagonal, a diagonal metric is consistent. Our metric ansatz is  \cite{Ammon:2009xh}
\begin{equation}
\label{eq:metricansatz}
\dd s^2 = -N(r)\sigma(r)^2\dd t^2 + \frac{1}{N(r)}\dd r^2 +r^2 f(r)^{-4}\dd x^2 + r^2f(r)^2\left(\dd y^2 + \dd z^2\right)\,,
\end{equation}
with $N(r)=-\frac{2m(r)}{r^2}+\frac{r^2}{L^2}$. For our black hole solutions we will denote the position of the horizon as $r_h$. The AdS boundary will be at $r\rightarrow\infty$. 

Inserting our ansatz into the Einstein and Yang-Mills equations yields five equations of motion for  $m(r),\,\sigma(r),\,f(r),\,\phi(r),\,w(r)$ and one constraint equation from the $rr$ component of the Einstein equations. The dynamical equations can be recast as (prime denotes $\frac{\partial}{\partial r}$)

\begin{equation}\label{eom}
 \begin{split}
m' &= \frac{\alpha^2 r f^4 w^2 \phi^2}{6 N \sigma^2} + \frac{\alpha^2 r^3 {\phi'}^2}{6 \sigma^2} + N\left(\frac{r^3{f'}^2}{f^2} + \frac{\alpha^2}{6} r f^4 {w'}^2\right) \,, \\ \sigma' &= \frac{\alpha^2 f^4 w^2 \phi^2}{3 r N^2 \sigma} + \sigma\left(\frac{2 r {f'}^2}{f^2} + \frac{\alpha^2 f^4 {w'}^2}{3 r}\right) \,, \\ f'' &= -\frac{\alpha^2 f^5 w^2 \phi^2}{3 r^2 N^2 \sigma^2} + \frac{\alpha^2 f^5 {w'}^2}{3 r^2} - f'\left(\frac{3}{r} - \frac{f'}{f} + \frac{N'}{N} +\frac{\sigma'}{\sigma}\right) \,, \\ \phi'' &= \frac{f^4 w^2 \phi}{r^2 N} - \phi'\left(\frac{3}{r} + \frac{\sigma'}{\sigma}\right) \,, \\ w'' &= -\frac{w \phi^2}{N^2 \sigma^2} - w'\left( \frac{1}{r} + \frac{4 f'}{f} + \frac{N'}{N} + \frac{\sigma'}{\sigma} \right).  \,
\end{split}
\end{equation}

The equations of motion are invariant under four scaling transformations (invariant quantities are not shown),
\begin{align*}
&(I)\quad && \sigma\rightarrow \lambda\sigma\,,\: &&  \phi\rightarrow \lambda\phi,&\\
&(II) && f\rightarrow \lambda f\,,  && w\rightarrow \lambda^{-2} w,&\\
&(III) && r\rightarrow \lambda r\,, && m\rightarrow \lambda^4 m \,,\: && w\rightarrow \lambda w\,,\: && \phi\rightarrow \lambda\phi\,, \: &&\\  
&(IV) && r\rightarrow \lambda r\,, && m\rightarrow \lambda^2 m\,,  && L\rightarrow \lambda L\,, &&\phi\rightarrow \frac{\phi}{\lambda}\,, && \alpha\rightarrow \lambda \alpha\,,
\end{align*}
where in each case $\lambda$ is some real positive number. Using (I) and (II) we can set the boundary values of both $\sigma(r)$ and $f(r)$ to one, so that the metric will be asymptotically AdS. We are free to use (III) to set $r_h$ to be one, but we will retain $r_h$ as a bookkeeping device. We will use (IV) to set the AdS radius $L$ to one.

A known analytic solution of the equations of motion is an asymptotically AdS Reissner-Nordstr\"om black hole, which has $\phi(r)=\mu - q/r^2$, $w(r)=0$, $\sigma(r)=f(r)=1$, and $N(r)= \left(r^2 - \frac{2m_0}{r^2} + \frac{2\alpha^2 q^2}{3 r^4}\right)$, where $m_0=\frac{r_h^4}{2}+\frac{\alpha^2 q^2}{3r_h^2}$ and $q= \mu r^2_h$. Here $\mu$ is the value of $\phi(r)$ at the boundary, which is the $U(1)_3$ chemical potential in dual field theory.

To find solutions with nonzero $w(r)$ we resort to numerics. We will solve the equations of motion using a shooting method. We will vary the values of functions at the horizon until we find solutions with suitable values at the AdS boundary. We thus need the asymptotic form of solutions both near the horizon $r=r_h$ and near the boundary $r=\infty$.

Near the horizon, we define $\epsilon_h\equiv\frac{r}{r_h}-1\ll 1$ and then expand every function in powers of $\epsilon_h$ with some constant coefficients. Two of these we can fix as follows. We determine $r_h$ by the condition $N(r_h)=0$, which gives that $m(r_h)=r_h^4/2$. Additionally, we must impose $A^3_t(r_h)=\phi(r_h)=0$ for $A$ to be well-defined as a one-form (see for example ref. \cite{Kobayashi:2006sb}). The equations of motion then impose relations among all the coefficients. A straightforward exercise shows that only four coefficients are independent,
\begin{equation}
\left\{\phi^h_1, \sigma^h_0, f^h_0, w^h_0\right\} \,,
\end{equation}
where the subscript denotes the order of $\epsilon_h$ (so $\sigma^h_0$ is the value of $\sigma(r)$ at the horizon, etc.). All other near-horizon coefficients are determined in terms of these four.

Near the boundary $r=\infty$ we define $\epsilon_b\equiv \left(\frac{r_h}{r}\right)^2\ll1$ and then expand every function in powers of $\epsilon_b$ with some constant coefficients. The equations of motion again impose relations among the coefficients. The independent coefficients are
\begin{equation}
\label{eq:coeffb}
\left\{m^b_0, \mu, \phi^b_1, w^b_1, f^b_2\right\} \,,
\end{equation}
where here the subscript denotes the power of $\epsilon_b$. All other near-boundary coefficients are determined in terms of these.

We used scaling symmetries to set $\sigma_0^b = f_0^b=1$. Our solutions will also have $w_0^b=0$ since we do not want to source the operator $\calj^x_1$ in the dual field theory ($U(1)_3$ will be \textit{spontaneously} broken). In our shooting method we choose a value of $\mu$ and then vary the four independent near-horizon coefficients until we find a solution which produces the desired value of $\mu$ and has $\sigma_0^b = f_0^b=1$ and $w_0^b=0$.

In what follows we will often work with dimensionless coefficients by scaling
out factors of $r_h$. We thus define the dimensionless functions
$\mt(r)\equiv m(r)/r_h^4$, $\tilde\phi(r)\equiv \phi(r)/r_h$ and $\tilde w(r)\equiv w(r)/r_h$,
while $f(r)$ and $\sigma(r)$ are already dimensionless.

\subsection{Thermodynamics}
\label{sec:Thermodynamics}
Next we will describe how to extract thermodynamic information from our solutions \cite{Ammon:2009xh}. Our solutions describe thermal equilibrium states in the dual field theory. We will work in the grand canonical ensemble, with fixed chemical potential $\mu$.

We can obtain the temperature and entropy from horizon data. The temperature $T$ is given by the Hawking temperature of the black hole,
\begin{equation}
  \label{eq:temperature}
  T=\frac{\kappa}{2\pi}=\frac{\sigma^h_0}{12\pi}\left(12-\alpha^2 \frac{{(\phit^h_1)}^2}{{\sigma^h_0}^2}\right)\,r_h\,.
\end{equation}
Here $\kappa=\left . \sqrt{\del_M \xi \del^M \xi} \right |_{r_h}$ is the surface gravity of the black hole, with $\xi$ being the norm of the timelike Killing vector, and in the second equality we write $T$ in terms of near-horizon coefficients. In what follows we will often convert from $r_h$ to $T$ simply by inverting the above equation. The entropy $S$ is given by the Bekenstein-Hawking entropy of the black hole,
\begin{equation}
  \label{eq:entropy}
  S=\frac{2\pi}{\k_5^2}A_h=\frac{2\pi V}{\k_5^2}r_h^3=  \frac{2\pi^4}{\k_5^2}VT^3
  \frac{12^3{\sigma_0^h}^3}{\left(12{\sigma_0^h}^2-{(\phit_1^h)}^2\alpha^2\right)^3}\,,
\end{equation}
where $A_h$ denotes the area of the horizon and $V = \int\!\dd^3x$.

The central quantity in the grand canonical ensemble is the grand potential $\Omega$. In AdS/CFT we identify $\Omega$ with $T$ times the on-shell bulk action in Euclidean signature. We thus analytically continue to Euclidean signature and compactify the time direction with period $1/T$. We denote the Euclidean bulk action as $I$ and $I_{\text{on-shell}}$ as its on-shell value (and similarly for other on-shell quantities). Our solutions will always be static, hence $I_{\text{on-shell}}$ will always include an integration over the time direction, producing a factor of $1/T$. To simplify expressions, we will define $I \equiv \tilde{I}/T$.  Starting now, we will refer to $\tilde{I}$ as the action. $\tilde{I}$ includes a bulk term, a Gibbons-Hawking boundary term, and counterterms,
\begin{equation}
\label{eq:renomaction}
  \tilde{I}=\tilde{I}_{\text{bulk}}+\tilde{I}_{\text{GH}}+\tilde{I}_{\text{CT}}\,.
\end{equation}
$\tilde{I}_{\text{bulk}}^{\text{on-shell}}$ and $\tilde{I}_{GH}^{\text{on-shell}}$ exhibit divergences, which are canceled by the counterterms in $\tilde{I}_{\text{CT}}$. To regulate these divergencies we introduce a hypersurface $r=r_{\text{bdy}}$ with some large but finite $r_{\text{bdy}}$. We will ultimately remove the regulator by taking $r_{\text{bdy}}\to\infty$. For our ansatz, the explicit form of  the three terms may be found in \cite{Ammon:2009xh}. Finally, $\Omega$ is related to the on-shell action, $\tilde{I}_{\text{on-shell}}$, as
\begin{equation}
\Omega = \tilde{I}_{\text{on-shell}}.
\end{equation}

The chemical potential $\mu$ is simply the boundary value of $A^3_t(r) = \phi(r)$. The charge density $\langle \calj^t_3\rangle$ of the dual field theory can be extracted from $\tilde{I}_{\text{on-shell}}$ by
\begin{equation}
\label{eq:density}
 \langle \calj^t_3\rangle= \lim_{r_{\text{bdy}}\rightarrow \infty} \frac{\delta \tilde{I}_{\text{on-shell}}}{\delta A_t^3(r_{\text{bdy}})} =-\frac{2\pi^3\alpha^2}{\k_5^2}T^3
 \frac{12^3{\sigma_0^h}^3}{\left(12{\sigma_0^h}^2-{(\phit_1^h)}^2\alpha^2\right)^3} \, \tilde{\phi}^b_1 \,.
\end{equation}
Similarly, the current density $\langle \calj^x_1\rangle$ is
\begin{equation}
\label{eq:condensate}
 \langle \calj^x_1\rangle= \lim_{r_{\text{bdy}}\rightarrow \infty} \frac{\delta \tilde{I}_{\text{on-shell}}}{\delta A_x^1(r_{\text{bdy}})} =+\frac{2\pi^3\alpha^2}{\k_5^2}T^3
 \frac{12^3{\sigma_0^h}^3}{\left(12{\sigma_0^h}^2-{(\phit_1^h)}^2\alpha^2\right)^3} \, \tilde{w}^b_1 \,.
\end{equation}

The expectation value of the stress-energy tensor of the CFT is \cite{Balasubramanian:1999re, deHaro:2000xn}
\begin{equation}
  \label{eq:energymombdy}
  \langle \calt_{\mu\nu}\rangle=\lim_{r_{\text{bdy}}\rightarrow \infty} \frac{2}{\sqrt{\gamma}}\frac{\delta \tilde{I}_{\text{on-shell}}}{\delta \gamma^{\mu\nu}}= \lim_{r_{\text{bdy}}\rightarrow \infty} \left [ \frac{r^2}{\k_5^2}
  \left(-K_{\mu\nu}+{K^\rho}_\rho\gamma_{\mu\nu}-3\,\gamma_{\mu\nu}\right) \right ]_{r=r_{\text{\text{bdy}}}} \,,
\end{equation}
where small Greek letter as indices run over the dual field theory directions $\{t,x,y,z\}$ and $K_{\mu\nu}= \frac{1}{2} \sqrt{N(r)} \, \partial_r \gamma_{\mu\nu}$ is the extrinsic curvature. We find
\begin{equation}
\label{eq:cftstressenergytensor}
 \begin{split}
\langle \calt_{tt} \rangle&=3\frac{\pi^4}{\k_5^2}T^4\frac{12^4{\sigma_0^h}^4}{\left(12{\sigma_0^h}^2-{(\phit_1^h)}^2\alpha^2\right)^4}\,\mt^b_0 \,,\\
  \langle \calt_{xx} \rangle&= \frac{\pi^4}{\k_5^2}T^4\frac{12^4{\sigma^h_0}^4}{\left(12{\sigma^h_0}^2-{(\phit_1^h)}^2\alpha^2\right)^4}\left(\mt^b_0-8f_2^b\right)\,,\\
  \langle \calt_{yy} \rangle = \langle \calt_{zz} \rangle&= \frac{\pi^4}{\k_5^2}T^4\frac{12^4{\sigma^h_0}^4}{\left(12{\sigma^h_0}^2-{(\phit_1^h)}^2\alpha^2\right)^4}\left(\mt^b_0+4f_2^b\right)\,.
 \end{split}
\end{equation}
Notice that $\langle \calt_{tx} \rangle = \langle \calt_{ty} \rangle = \langle \calt_{tz}
\rangle = 0$. Even in phases where the current $\langle \calj_1^x \rangle$ is
nonzero, the fluid will have zero net momentum. Indeed, this result is
guaranteed by our ansatz for the gauge field which implies a diagonal
Yang-Mills stress-energy tensor and a diagonal metric. The spacetime is static.

Tracelessness of the stress-energy tensor (in Lorentzian signature) implies $\langle \calt_{tt} \rangle= \langle \calt_{xx} \rangle + \langle \calt_{yy} \rangle + \langle \calt_{zz} \rangle$, which is indeed true for eq. (\ref{eq:cftstressenergytensor}), so in the dual field theory we always have a conformal fluid. The only physical parameter in the dual field theory is thus the ratio $\mu/T$.

For $\mt^b_0=\frac12+\frac{\alpha^2 \tilde\mu^2}{3}$, $\sigma_0^h=1$, ${\phit_1^h}=2\tilde\mu$, $f_2^b=0$, and $\tilde\phi^b_1=-\tilde\mu$ we recover the correct thermodynamic properties of the Reissner-Nordstr\"om black hole, which preserves the $SO(3)$ rotational symmetry. For example, we find that $\langle \calt_{xx} \rangle=\langle \calt_{yy} \rangle = \langle \calt_{zz} \rangle$ and $\Omega=-V\langle \calt_{yy}\rangle$, \ie $\langle \calt_{yy}\rangle$ is the pressure $P$. For solutions with nonzero $\langle \calj_1^x\rangle$, the $SO(3)$  is broken to $SO(2)$. In these cases, we find that $\langle \calt_{xx} \rangle \neq \langle \calt_{yy} \rangle = \langle \calt_{zz} \rangle$. In the superfluid phase, both the nonzero $\langle \calj^x_1 \rangle$ and the stress-energy tensor indicate breaking of $SO(3)$.  Just using the equations above, we also find
\begin{equation}
\label{eq:grandpot}
	\Omega = - V\langle \calt_{yy} \rangle\,.
\end{equation}
This again suggest the identification of $\langle \calt_{yy}\rangle$ as the pressure $P$. However due to the breaking of the $SO(3)$ symmetry $\langle \calt_{xx}\rangle$ is not the pressure $P$ but most also contain terms which are non-zero in the broken phase, \ie terms which contain the order parameter $\langle \calj^x_1\rangle$. For instance it may be written as
\begin{equation}
\label{eq:stressenergyxx}
\langle \calt_{xx}\rangle=P+\Delta\, \langle \calj^x_1\rangle \langle \calj^x_1\rangle\,,
\end{equation}
where $\Delta$ is a measure for the breaking of the rotational symmetry and is given by
\begin{equation}
\label{eq:Delta}
\Delta=-\frac{3\k_5^2}{\alpha^2\pi^2T^2}\frac{\left(12{\sigma_0^h}^2-{(\phit_1^h)}^2\alpha^2\right)^2}{12^2{\sigma_0^h}^2}\frac{f_2^b}{\left(\tilde w_1^b\right)^2}\,.
\end{equation}

Using this identification we can write down the stress-energy tensor for the dual field theory in equilibrium in a covariant form
\begin{equation}
\label{eq:stressenergycovariant}
\langle \calt^{\mu\nu}\rangle=\epsilon u^\mu u^\nu+P\,P^{\mu\nu}+\Delta\, {P^\mu}_\lambda{P^\nu}_\rho \langle \calj^\lambda_a\rangle \langle \calj^\rho_a\rangle\,,
\end{equation}
where $\epsilon=\langle \calt^{tt}\rangle$ is the energy density and $P^{\mu\nu}=u^\mu u^\nu+\eta^{\mu\nu}$ is the projector to the space perpendicular to the velocity $u^\mu$.

\begin{figure}[t]
\centering
\subfigure[]{\includegraphics[width=0.48\textwidth]{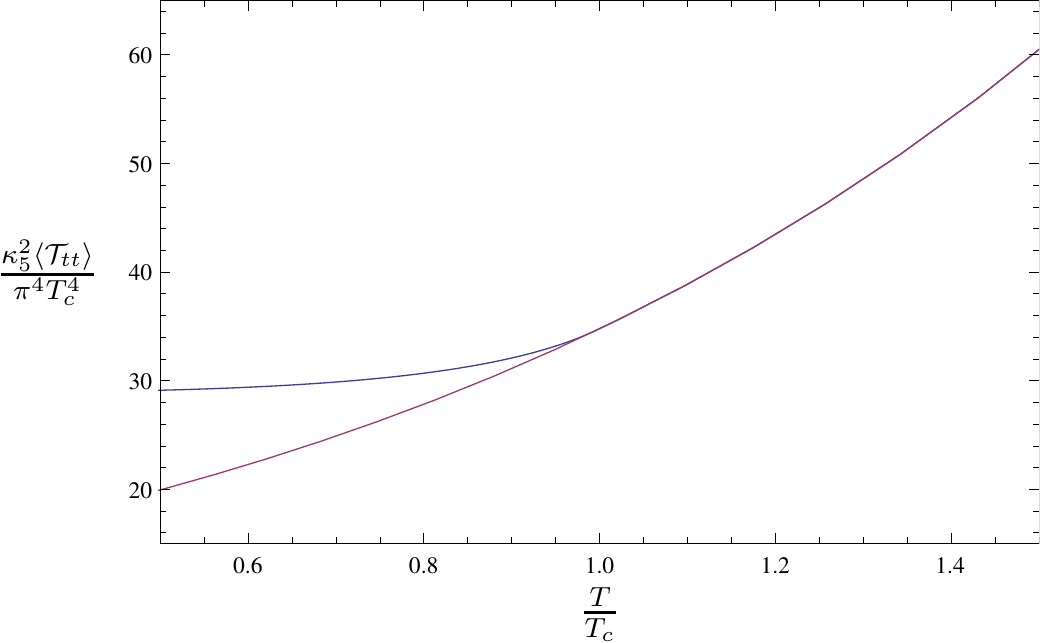}}
\hfill
\subfigure[]{\includegraphics[width=0.48\textwidth]{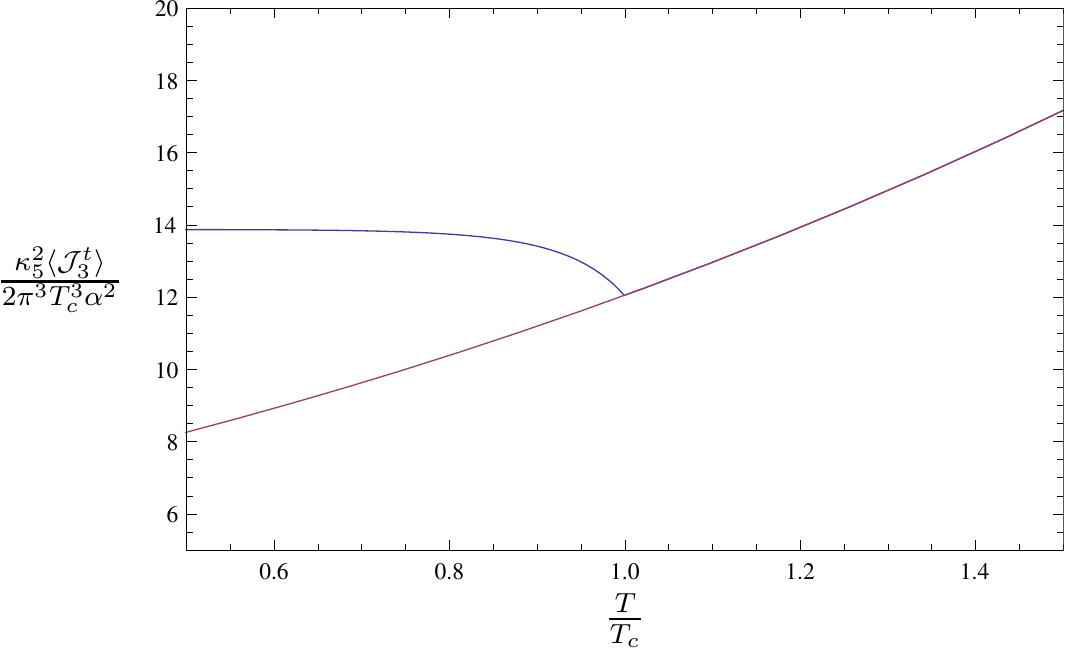}}
\caption{The energy density $\langle \calt_{tt} \rangle$ (a) and the charge density $\langle \calj^t_3 \rangle$ (b) over the reduced temperature $T/T_c$ for $\alpha=0.316$. The red line is the solution without a condensate and the blue line the solution with $\langle \calj^1_x\rangle \neq 0$ below $T_c$.}
\label{fig:Ttt}
\end{figure}

In figure \ref{fig:Ttt} we plot $\langle \calt_{tt}\rangle$ and $\langle \calj^t_3\rangle$ versus the reduced temperature, respectively. We see that in both cases there is one solution for temperatures above $T_c$ and two for temperatures below $T_c$. From considerations in \cite{Ammon:2009xh} we know that the solution with condensate (blue line) is the thermodynamically preferred one. For further plots see \cite{Ammon:2009xh}.

\begin{figure}[t]
\centering
\includegraphics[width=0.8\textwidth]{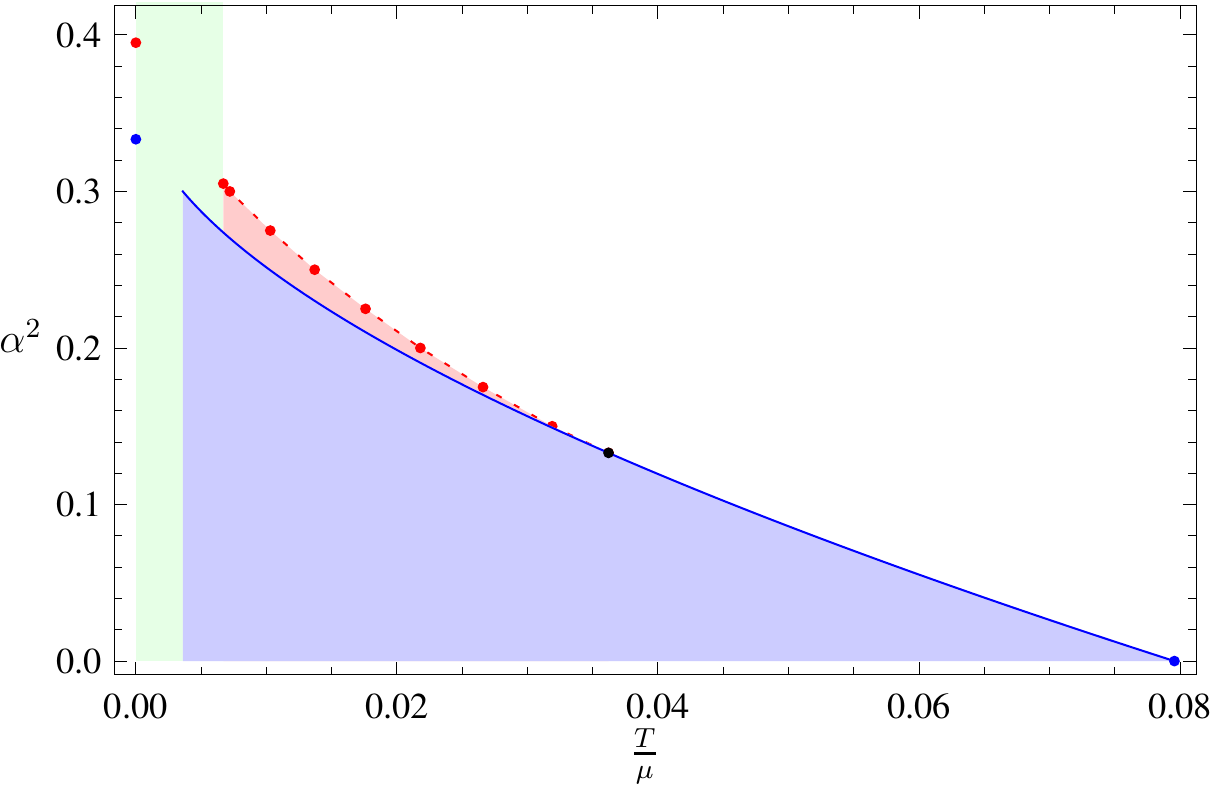}
\caption{The phase structure of the theory: In the blue and red region the broken phase is the thermodynamically preferred phase while in the white region the Reissner-Nordstr\"om black hole is the ground state. In the blue region the Reissner-Nordstr\"om black hole is unstable and the transition from the white to the blue region is second order. In the red region the Reissner-Nordstr\"om black hole is still stable. The transition form the white to the red region is first order. The black dot determines the critical point where the order of the phase transition changes. In the green region we cannot trust our numerics.}
\label{fig:phasediag}
\end{figure}

In addition in \cite{Ammon:2009xh} it was found that the order of the phase transition depends on the ratio of the coupling constants $\alpha$. For $\alpha\le \alpha_c=0.365$, the phase transition is second order while for larger values of $\alpha$ the transition becomes first order. The critical temperature decreases as we increase the parameter $\alpha$. The quantitative dependence of the critical temperature on the parameter $\alpha$ is given in figure~\ref{fig:phasediag}. The broken phase is thermodynamically preferred in the blue and red region while in the white region the Reissner-Nordstr\"om black hole is favored. The Reissner-Nordstr\"om black hole is unstable in the blue region and the phase transition from the white to the blue region is second order. In the red region, the Reissner-Nordstr\"om black hole is still stable however the state with non-zero condensate is preferred. The transition from the white to the red region is first order. In the green region we cannot trust our numerics. At zero temperature, the data is obtained as described in \cite{Basu:2009vv,Gubser:2010dm}.


\section{Perturbations about Equilibrium}
\label{sec:Perturbations-about-the}
In this section we study the response of the holographic p-wave superfluid under small perturbations. On the gravity side these perturbations are given by fluctuations of the metric $h_{MN}(x^\mu,r)$ and the gauge field $a^a_{M}(x^\mu,r)$. Thus we study in total $14$ physical modes: $5$ from the massless graviton in 5 dimensions and  $3\times 3$ from the massless vectors in five dimensions.  Due to time and spatial translation invariance in the Minkowski directions, the fluctuations can be decomposed in a Fourier decomposition
\begin{equation}
\label{eq:fourier}
\begin{split}
h_{MN}(x^\mu,r)&=\int\!\frac{\dd^4k}{(2\pi)^4}\ee^{\ii k_\mu x^\mu}\,\hat{h}_{MN}(k^\mu,r)\,,\\
a^a_{M}(x^\mu,r)&=\int\!\frac{\dd^4k}{(2\pi)^4}\ee^{\ii k_\mu x^\mu}\,\hat{a}^a_{M}(k^\mu,r)\,.
\end{split}
\end{equation}
To simplify notations we drop the hat on the transformed fields which we use from now on if not stated otherwise.

\subsection{Characterization of Fluctuations and Gauge Fixing}
\label{sec:Characterization-of-Fluctuations}
In general we have to introduce two spatial momenta: one longitudinal to the condensate $k_{\|}$ and one perpendicular to the condensate $k_\perp$, \ie $k^\mu=(\omega,k_\|,k_\perp,0)$. Introducing the momentum perpendicular to condensate breaks the remaining rotational symmetry $SO(2)$ down the discrete $\mathbb{Z}_2$ parity transformation $P_\perp$: $k_\perp\to -k_\perp$ and $x_\perp\to -x_\perp$. Thus introducing this momentum forbids the usual classification of the fluctuations in different helicity states of the little group since the symmetry group just consists of discrete groups at best $P_\|\times P_\perp$. We do not study this case further in this paper. However a momentum exclusively in the direction longitudinal to the condensate or zero spatial momentum preserves the $SO(2)$ rotational symmetry such that we can classify the fluctuations according to their transformation under this $SO(2)$ symmetry (see table~\ref{tab:classification}). The modes of different helicity decouple from each other. The momentum longitudinal to the condensate, however, breaks the longitudinal parity invariance $P_\|$.

\begin{table}
\centering
\begin{tabular}{l|ccc}
& dynamical fields & constraints &  \# physical modes\\
\hline
helicity 2 & $h_{yz},h_{yy}-h_{zz}$ & none & 2\\
helicity 1 & $h_{ty},h_{xy};a^a_{y}$ & $h_{yr}$ & 4\\
	        & $h_{tz},h_{xz};a^a_{z}$ & $h_{zr}$ & 4\\
helicity 0 & $h_{tt},h_{xx},h_{yy}+h_{zz},h_{xt};a^a_t,a^a_x$ & $h_{tr},h_{xr},h_{rr};a^a_r$ & 4
\end{tabular}
\caption{Classifications of the fluctuations according to their transformation under the little group $SO(2)$. The constraints are given by the equations of motion for the fields which are  set to zero due the fixing of the gauge freedom:  $a_r^a\equiv 0$ and $h_{rM}\equiv 0$. The number of physical modes is obtained by the number of dynamical fields minus the number of constraints. Due to $SO(2)$ invariance the fields in the first and second line of the helicity one fields can be identified.}
\label{tab:classification}
\end{table}

In order to obtain the physical modes of the system we have to fix the gauge freedom. We choose a gauge where $a_r^a\equiv 0$ and $h_{Mr}\equiv 0$ such that the equations of motion for these fields become constraints. These constraints fix the unphysical fluctuations in each helicity sector and allow only the physical modes to fluctuate. The physical modes may be constructed by enforcing them to be invariant under the residual gauge transformations, $\delta a_r^a=0$ and $\delta h_{Mr}=0$ (see appendix~\ref{sec:Constructing-the-Gauge}),
\begin{equation}
\label{eq:physicalmodes}
\begin{aligned}
&\text{helicity two:}\; && \Xi=g^{yy}h_{yz},h_{yy}-h_{zz}\,,\\
&\text{helicity one:} && \Psi=g^{yy}(\omega h_{xy}+k_\| h_{ty}); a^a_y\,,
\end{aligned}
\end{equation}
and helicity zero:
\begin{equation}
\label{eq:physicalmodeshelicityzero}
     \begin{split}
      \Phi_1 =& \xi_y,\\
      \Phi_2 =&a^1_t+\frac{i \omega}{\phi}a^2_t+\frac{i k \left(\omega ^2-\phi^2\right)}{\left(k^2-w^2\right) \phi}a^2_x+\frac{w \left(\omega ^2-\phi^2\right)}{\left(k^2-w^2\right) \phi}a^3_x,\\
      \Phi_3 =&\xi_x-\frac{k^2 {N\sigma^2f^4}}{\omega ^2 {r^2}}\xi_t+\frac{2 k}{\omega}\xi_{tx},\\
      \Phi_4 =&a^1_x+\frac{k}{\omega}a^1_t-\frac{1}{2} w \xi_x-\frac{w'}{\phi'}a^3_t+\frac{\phi w'}{2 \phi'}\xi_t-\frac{k \left(\omega ^2 w'+w \phi \phi'\right)}{\omega  \left(k^2-w^2\right) \phi'}a^3_x\\&-\ii\frac{\omega^2ww'+k^2\phi\phi'}{\omega\phi'(k^2-w^2)}a^2_x ,
     \end{split}
    \end{equation}
with
\begin{equation}
\label{eq:defxi}
  \xi_y = g^{yy}h_{yy}, \qquad \xi_x = g^{xx}h_{xx}, \qquad \xi_t = g^{tt}h_{tt}, \qquad \xi_{tx} = g^{xx}h_{tx}.
\end{equation}

\subsection{Equations of Motion, On-shell Action and Correlators}
\label{sec:Equation-of-Motion}
In the following we will focus on the response exclusively due to time dependent perturbations, \ie $k^\mu=(\omega,0,0,0)$. In this case in addition to the $SO(2)$ symmetry, $P_\|$ parity is conserved which allows us to decouple some of the physical modes in the different helicity blocks. In this section we write down the equations of motion for the fluctuations, determine the on-shell action and vary the on-shell action with respect to the fluctuations to obtain the retarded Green's functions $G$ of the stress-energy tensor $T^{\mu\nu}$ and the currents $J^\mu_a$,
\begin{equation}
\label{eq:correlators}
\begin{split}
G^{\mu\nu,\rho\sigma}(k)&=-\ii\int\!\dd t \dd^3 x\,\ee^{-\ii k x}\;\theta(t)\langle[T^{\mu\nu}(t,\vec{x}),T^{\rho\sigma}(0,0)]\rangle\,,\\
G_{a,b}^{\mu,\nu}(k)&=-\ii\int\!\dd t\dd^3 x\,\ee^{-\ii k x}\;\theta(t)\langle[J_a^\mu(t,\vec{x}),J_b^\nu(0,0)]\rangle\,,\\
{G^{\mu\nu}}^\rho_a(k)&=-\ii\int\!\dd t\dd^3 x\,\ee^{-\ii k x}\;\theta(t)\langle[T^{\mu\nu}(t,\vec{x}),J_a^\rho(0,0)]\rangle\,,\\
{G_a^\rho}^{\mu\nu}(k)&=-\ii\int\!\dd t\dd^3 x\,\ee^{-\ii k x}\;\theta(t)\langle[J_a^\rho(t,\vec{x}),T^{\mu\nu}(0,0)]\rangle\,.
\end{split}
\end{equation}
$T^{\mu\nu}$ and $J^\mu_a$ are the full stress-energy tensor and current, respectively. Thus they include the equilibrium parts, $\langle \calt^{\mu\nu}\rangle$ and $\langle \calj^\mu_a\rangle$, as well as the corresponding dissipative parts which arise due to the inclusion of fluctuations in our model. In the following we split the analysis into the different helicity blocks.

\subsubsection{Helicity two mode}
\label{sec:Helicity-two-mode}
First we look at the non-trivial helicity two mode displayed in table \ref{tab:classification}. If we expand the action \eqref{eq:action} up to second order in the fluctuations, this mode decouples from every other field. Therefore it can be written as a minimal coupled scalar with the equation of motion
\begin{equation}
\label{eq:hel2eom}
\Xi'' +\left(\frac{1}{r}+\frac{4 r}{N} - \frac{r \alpha ^2 {\phi'}^2}{3 N \sigma^2}\right) \Xi'+\frac{\omega ^2}{N^2 \sigma^2}\Xi=0\,.
\end{equation}
The contribution from this mode to the on-shell action is
\begin{equation}
\label{eq:hel2action}
\tilde S^{\text{on-shell}}_{\text{helicity 2}}=\frac{1}{\k_5^2}\int\!\frac{\dd^4 k}{(2\pi)^4}\left\{r^3N\sigma\left[\left(\frac{3}{2\sqrt{N}}-\frac{1}{r}+\frac{f'}{2 f}-\frac{N'}{4N}-\frac{\sigma'}{2\sigma}\right)\Xi^2-\frac{1}{4} \Xi \Xi'\right]\right\}_{r=r_\text{bdy}}\,,
\end{equation}
which is divergent as we send $r_{\text{bdy}}\to\infty$. The divergence can be cured by holographic renormalization (see appendix~\ref{sec:Holographic-Renormalization}). The renormalized on-shell action is
\begin{equation}
\label{eq:hel2actionren}
S^{\text{on-shell}}_{\text{helicity 2}}=\frac{r_h^4}{\k_5^2}\int\!\frac{\dd^4 k}{(2\pi)^4}\left[\Xi^b_0\Xi^b_2-\frac{1}{2}\left(\tilde m_0^b+4 f_2^b-\frac{1}{32}\tilde \omega^4\right) \left(\Xi_0^b\right)^2\right]\,,
\end{equation}
where $\tilde\omega=\omega/r_h$ is the dimensionless frequency, $\Xi^b_0$ and $\Xi^b_2$ are defined similarly to the quantities in \eqref{eq:coeffb} and $\Xi_0^b\Xi_2^b$ is a short form for $\Xi_0^b(\omega)\Xi_2^b(-\omega)$. Now we use the recipe by Son and Starinets \cite{Son:2002sd} to compute the Green's function of this component. The response due to the perturbation $h_{yz}$ is given by
\begin{equation}
\label{eq:hel2responsea}
\langle T^{yz}\rangle(\omega)=\frac{\delta S^{\text{on-shell}}_{\text{helicity 2}}}{\delta \Xi_0^b(-\omega)}=G^{yz,yz}(\omega)\Xi_0^b(\omega)\,,
\end{equation}
with
\begin{equation}
\label{eq:hel2response}
G^{yz,yz}(\omega)=\left(\frac{2r_h^4}{\k_5^2}\frac{\Xi_2^b(\omega)}{\Xi_0^b(\omega)}-\langle \calt_{yy}\rangle+\frac{1}{32}\omega^4\right)\,,
\end{equation}
where $\langle \calt_{yy}\rangle$ is the equilibrium contribution given by the pressure $P$. As we will see in section \ref{sec:Universal-Shear-Viscosity}, the Green's function of this helicity mode will lead to a shear viscosity component with universal behavior, i.e. $\eta_{yz}/s=1/4\pi$.

\subsubsection{Helicity one modes}
\label{sec:Helicity-one-modes}
Now we look at the helicity one modes displayed in table \ref{tab:classification}. Again we obtain their equations of motion by expanding the action \eqref{eq:action} up to second order in the fluctuations and varying it with respect to the corresponding fields. The equations of motion are
\begin{subequations}
\label{eq:hel1eomb1}
\begin{align}
0&={a^3_y}'' + \left(\frac{1}{r}-\frac{2 f'}{f}+\frac{N'}{N}+\frac{\sigma '}{\sigma }\right){a^3_y}'+ \left(\frac{\omega ^2}{N^2 \sigma ^2}-\frac{f^4 w^2}{r^2 N}-\frac{2 \alpha ^2 \phi '^2}{N\sigma ^2}\right){a^3_y}\,, \label{eq:eompwavefluca3y}\\
0&=\Psi_t'+\frac{2 \alpha ^2 \phi '}{r^2 f^2}{a^3_y}\,,\label{eq:eompwavefluca3ypsit}
\end{align}
\end{subequations}
and
\begin{subequations}
\label{eq:hel1eomb2}
\begin{align}
0=&\Psi_x''+\left(\frac{1}{r}+\frac{4 r}{N}+\frac{6 f'}{f}-\frac{r \alpha ^2 \phi '^2}{3 N \sigma ^2}\right)\Psi_x'+\frac{2 \alpha ^2 w'}{r^2 f^2}{a^1_y}'+\frac{\omega ^2}{N^2 \sigma ^2}\Psi_x\nonumber\\
       &+\frac{2 i \alpha ^2 \omega w \phi }{r^2 f^2 N^2 \sigma ^2}{a^2_y}-\frac{2 \alpha ^2 w \phi ^2}{r^2 f^2 N^2 \sigma ^2}{a^1_y}\,, \label{eq:eompwavefluchxy}\\
0=&{a^1_y}''+\left(\frac{1}{r}-\frac{2 f'}{f}+\frac{N'}{N}+\frac{\sigma '}{\sigma }\right){a^1_y}' -f^6 w' \Psi_x'+\left(\frac{\omega ^2}{N^2 \sigma ^2}+\frac{\phi ^2}{N^2 \sigma ^2}\right){a^1_y} \nonumber\\
	&-\frac{2 i \omega  \phi }{N^2 \sigma ^2}{a^2_y},\\
0=&{a^2_y}''+ \left(\frac{1}{r}-\frac{2 f'}{f}+\frac{N'}{N}+\frac{\sigma '}{\sigma }\right){a^2_y}'+ \left(\frac{\omega ^2}{N^2 \sigma ^2}+\frac{\phi ^2}{N^2 \sigma ^2}-\frac{f^4 w^2}{r^2 N}\right){a^2_y}\nonumber\\
	&+\frac{2 i \omega  \phi }{N^2 \sigma ^2}{a^1_y}-\frac{i \omega  f^6 w \phi}{N^2 \sigma ^2}\Psi_x.
\end{align}
\end{subequations}
where $\Psi_t=g^{yy}h_{ty}$ and $\Psi_x=g^{yy}h_{xy}$. Note that due to the parity $P_\|$, the helicity one modes split into two blocks where the modes of the first block are even while the modes of the second block are odd under $P_\|$. In the first block there is only one physical mode $a_y^3$ while the value of the other field $\Psi_t$ is given by the constraint \eqref{eq:eompwavefluca3ypsit}. This can also be seen in the gauge invariant fields \eqref{eq:physicalmodes} since $h_{ty}$ drop out for $k_\|=0$. The other three physical modes appear in the second block where $\Psi_x=\Psi$ for $k_\|=0$.

The contribution from these modes to the on-shell action is
\begin{equation}
\label{eq:hel1action}
\begin{split}
\tilde S^\text{on-shell}_{\text{helicity 1}} =& \frac{1}{\kappa_5^2}\int \frac{\dd^4 k}{(2\pi)^4}\ \bigg\{\frac{r^5 f^2}{4 \sigma}{\Psi_t} {\Psi_t}'-\frac{1}{4} r^3 f^6 N \sigma {\Psi_x} {\Psi_x}'-\frac{r \alpha ^2 N \sigma}{2 f^2}\left({a^1_y}{a^1_y}'+{a^2_y}{a^2_y}'+{a^3_y}{a^3_y}'\right)\\
&+\frac{3r^4f^2}{2\sigma}\left(1-\frac{r}{\sqrt{N}}\right) {\Psi_t}^2+\frac{r^3f^6N\sigma}{2}\left(\frac{3}{\sqrt{N}}-\frac{2}{r}-\frac{2f'}{f}-\frac{N'}{2N}-\frac{\sigma'}{\sigma}\right){\Psi_x}^2\\
&+\frac{r \alpha ^2 f^4 N \sigma w'}{2} {a^1_y}{\Psi_x}-\frac{r^3 \alpha ^2 \phi'}{2 \sigma}{a^3_y}{\Psi_t}
\bigg\}\bigg|_{r=r_\text{bdy}}\,, 
\end{split}
\end{equation}
which is again divergent\footnote{Note that the contribution of the on-shell action is zero at the horizon since we can set $\Psi_t$ to zero there.}. The renormalized on-shell action is given by
\begin{equation}
\label{eq:hel1actionren}
\begin{split}
S^\text{on-shell}_{\text{helicity 1}} =& \frac{r_h^4}{\kappa_5^2}\int \frac{\dd^4 k}{(2\pi)^4}\ \bigg\{{\big(\Psi_x\big)_0^b}{\big(\Psi_x\big)_2^b}+\alpha^2\left[{\left(\tilde a_y^1\right)_0^b}{\left(\tilde a_y^1\right)_1^b}+{\left(\tilde a_y^2\right)_0^b}{\left(\tilde a_y^2\right)_1^b}+{\left(\tilde a_y^3\right)_0^b}{\left(\tilde a_y^3\right)_1^b}\right]\\
&-\frac{1}{2}\left(\tilde m_0^b-8f_4^b-\frac{1}{32}\tilde \omega^4\right) \left({\big(\Psi_x\big)_0^b}\right)^2-\frac{3}{2}\tilde m_0^B \left({\big(\Psi_t\big)_0^b}\right)^2-\frac{1}{4} \alpha ^2 \tilde\omega ^2{\left(\tilde a_y^3\right)_0^b}^2\\
&-\frac{1}{4} \alpha ^2 (\tilde \mu ^2 + \tilde \omega^2)\left[{\left(\tilde a_y^1\right)_0^b}^2 + {\left(\tilde a_y^2\right)_0^b}^2\right]+\ii\alpha^2\tilde \omega\tilde\mu {\left(\tilde a_y^1\right)_0^b} {\left(\tilde a_y^2\right)_0^b}\\
&+\alpha^2\left[2{\tilde\phi_1^b}{\left(\tilde a_y^3\right)_0^b}{\big(\Psi_t\big)_0^b}-\tilde w_1^b{\left(\tilde a_y^1\right)_0^b}{\big(\Psi_x\big)_0^b}\right]\bigg\},
\end{split}
\end{equation}
where $\tilde a^a_\mu=r_h a^a_\mu$ is dimensionless. We obtain the response of the system due to the fluctuations $a_y^3$ and $h_{ty}$ by variation of the on-shell action,
\begin{equation}
\label{eq:hel1responseb1}
\begin{pmatrix}\langle J^y_3 \rangle(\omega)\\[1ex]  \langle T^{ty} \rangle(\omega) \end{pmatrix}=\begin{pmatrix}\frac{\delta S^\text{on-shell}_{\text{helicity 1}}}{\delta {\left(a_y^3\right)_0^b}(-\omega)}\\[2ex] \frac{\delta S^\text{on-shell}_{\text{helicity 1}}}{\delta {\big(\Psi_t\big)_0^b}(-\omega)}\end{pmatrix}=\begin{pmatrix}G_{3,3}^{y,y}(\omega) & {G_3^y}^{ty}(\omega)\\[1ex] {G^{ty}}^y_3(\omega) & G^{ty,ty}(\omega)\end{pmatrix}\begin{pmatrix}{\left(a_y^3\right)_0^b}(\omega)\\[1ex] {\big(\Psi_t\big)_0^b}(\omega)\end{pmatrix}\,,
\end{equation}
with
\begin{equation}
\label{eq:hel1corb1}
\begin{pmatrix}G_{3,3}^{y,y}(\omega) & {G_3^y}^{ty}(\omega)\\[1ex] {G^{ty}}^y_3(\omega) & G^{ty,ty}(\omega)\end{pmatrix}=\begin{pmatrix} \frac{\alpha^2r_h^2}{\kappa_5^2}\left(\frac{2{\left(\tilde a_y^3\right)_1^b}(\omega)}{{\left(\tilde a_y^3\right)_0^b}(\omega)} -\frac{\tilde\omega^2}{2}\right)\quad & - \langle \calj_3^t\rangle\\[1ex]  -\langle \calj_3^t\rangle &-\langle \calt_{tt}\rangle \end{pmatrix}.
\end{equation}
This result agrees with the result obtain in the holographic s-wave superfluids \cite{Hartnoll:2008kx,Hartnoll:2009sz} and thus the breaking of the rotational symmetry has no effect on this subset of fluctuations. The coupling between the current $J^y_3$ and the momentum $T_{ty}$ is known as the thermoelectric effect which we will study in the next section.

The response due to the fluctuations $a_y^1$, $a_y^2$ and $h_{xy}$ is given by
\begin{equation}
\label{eq:hel1responseb2}
\begin{pmatrix}\langle J^{y}_{1} \rangle(\omega)\\[1ex] \langle J^{y}_{2} \rangle(\omega)\\[1ex] \langle T^{xy} \rangle(\omega)\end{pmatrix}=\begin{pmatrix}\frac{\delta S^\text{on-shell}_{\text{helicity 1}}}{\delta {\left(a_y^1\right)_0^b}(-\omega)}\\[2ex] \frac{\delta S^\text{on-shell}_{\text{helicity 1}}}{\delta {\left(a_y^2\right)_0^b}(-\omega)}\\[2ex] \frac{\delta S^\text{on-shell}_{\text{helicity 1}}}{\delta {\big(\Psi_x\big)_0^b}(-\omega)}\end{pmatrix}=\begin{pmatrix}G_{1,1}^{y,y}(\omega) & G_{1,2}^{y,y}(\omega) & {G_1^y}^{xy}(\omega)\\[1ex] G_{2,1}^{y,y}(\omega) & G_{2,2}^{y,y}(\omega) & {G_2^y}^{xy}(\omega)\\[1ex] {G^{xy}}_1^y(\omega) & {G^{xy}}_2^y(\omega) & G^{xy,xy}(\omega)\end{pmatrix}
\begin{pmatrix} {\left(a_y^1\right)_0^b}(\omega)\\[1ex] {\left(a_y^2\right)_0^b}(\omega)\\[1ex]  {\big(\Psi_x\big)_0^b}(\omega)\end{pmatrix}\,,
\end{equation}
where the matrix of the Green's functions is given by
\begin{equation}
\begin{pmatrix} \frac{\alpha^2r_h^2}{\kappa_5^2}\left(2\frac{{\left(\tilde a_y^1\right)_1^b}(\omega)}{{\left(\tilde a_y^1\right)_0^b}(\omega)}-\frac{\tilde\mu^2+\tilde\omega^2}{2}\right) & \frac{\alpha^2r_h^2}{\kappa_5^2}\left(2\frac{{\left(\tilde a_y^1\right)_1^b}(\omega)}{{\left(\tilde a_y^2\right)_0^b}(\omega)}+\ii\tilde\omega\tilde\mu\right) &  -\frac{\langle \calj_1^x\rangle}{2} + 2\frac{\alpha^2r_h^3}{\kappa_5^2}\frac{{\left(\tilde a_y^1\right)_1^b}(\omega)}{{\big(\Psi_x\big)_0^b}(\omega)}\\[1ex] \frac{\alpha^2r_h^2}{\kappa_5^2}\left(2\frac{{\left(\tilde a_y^2\right)_1^b}(\omega)}{{\left(\tilde a_y^1\right)_0^b}(\omega)}-\ii\tilde\omega\tilde\mu\right) & \frac{\alpha^2r_h^2}{\kappa_5^2}\left(2\frac{{\left(\tilde a_y^2\right)_1^b}(\omega)}{{\left(\tilde a_y^2\right)_0^b}(\omega)} -\frac{\tilde\mu^2+\tilde\omega^2}{2}\right) &  2\frac{\alpha^2r_h^3}{\kappa_5^2}\frac{{\left(\tilde a_y^2\right)_1^b}(\omega)}{{\big(\Psi_x\big)_0^b}(\omega)}\\[1ex] -\frac{\langle \calj_1^x\rangle}{2} + 2\frac{r_h^3}{\kappa_5^2}\frac{{\big(\Psi_x\big)_2^b}(\omega)}{{\left(\tilde a_y^1\right)_0^b}(\omega)}& 2\frac{r_h^3}{\kappa_5^2}\frac{{\big(\Psi_x\big)_2^b}(\omega)}{{\left(\tilde a_y^2\right)_0^b}(\omega)} & \frac{r_h^4}{\kappa_5^2}\left(2\frac{{\big(\Psi_x\big)_2^b}(\omega)}{{\big(\Psi_x\big)_0^b}(\omega)}+\frac{1}{32}\tilde\omega^4\right)-\langle \calt_{xx}\rangle \end{pmatrix}\,.
\end{equation}
Due to the breaking of the rotational symmetry we see a new coupling between the currents $J_{1,2}^y$ and the stress tensor $T_{xy}$ in this subset of the fluctuations. This new coupling generates some interesting new physical effect: it induces a non-universal behavior of the ratio of shear viscosity to entropy density and a flexoelectric effect known from nematic crystals. We will study these effects in the next section.


\section{Transport Properties}
\label{sec:Transport-Properties}
In this section we extract the transport properties of the holographic p-wave superfluid from the correlation functions presented in the previous section. We split our analysis into distinct transport phenomena.

\subsection{Universal Shear Viscosity}
\label{sec:Universal-Shear-Viscosity}
Let us start by considering the helicity two mode $h_{yz}$. It is well known that, in the isotropic case, the corresponding component of the energy-momentum tensor may be written as\footnote{Note that $g^{\mu\nu}=\eta^{\mu\nu}-h^{\mu\nu}$.}
\begin{equation}
\langle T^{yz}\rangle = -(P+\ii\omega\eta_{yz}) h_{yz}\,.
\end{equation}
Using ~\eqref{eq:off-diagonalstressanisotroptic} we see that this result is still correct in the transversely symmetric case we are studying here. The result also agrees with our gravity calculation ~\eqref{eq:hel2response} \footnote{We do not see an $\omega^4$-term as in \eqref{eq:hel2response} in the linear hydrodynamic description since this term corresponds to higher order term with four derivatives.}
. Thus the shear viscosity is given by the well-known Kubo formula
\begin{equation}
\label{eq:dereta}
\eta_{yz} =-\lim_{\omega\to0}\frac{1}{\omega}\im\left(G^{yz,yz}\right)=-\lim_{\omega\to0}\frac{1}{\omega}\frac{2r_h^4}{\kappa_5^2}\frac{\Xi_2^b(\omega)}{\Xi_0^b(\omega)}\,.
\end{equation}
In the following we show that we can apply the proof for the universal result for the ratio of the shear viscosity to entropy density described in \cite{Iqbal:2008by}. In the $\omega\to0$ limit the equation of motion \eqref{eq:hel2eom} corresponds to $\partial_r \Pi = 0$, with $\Pi$ the conjugate momentum to the field $\Xi=g^{yy} h_{yz}$. This is the decisive condition in order to apply the proof of \cite{Iqbal:2008by}. Therefore we conclude that here we obtain the universal result for the ratio of shear viscosity to entropy density \cite{Kovtun:2004de,Buchel:2003tz,Iqbal:2008by},
\begin{equation}
\label{eq:universalshear}
\frac{\eta_{yz}}{s}=\frac{1}{4\pi}\,.
\end{equation}

In this subset we do not see any effect of the rotational symmetry breaking since the fluctuation $h_{yz}$ is transverse to the condensate.
\subsection{Thermoelectric Effect perpendicular to the Condensate}
\label{sec:Thermoelectric-Effect-perpendicular}
\begin{figure}[b]
\centering
\includegraphics[width=0.9\linewidth]{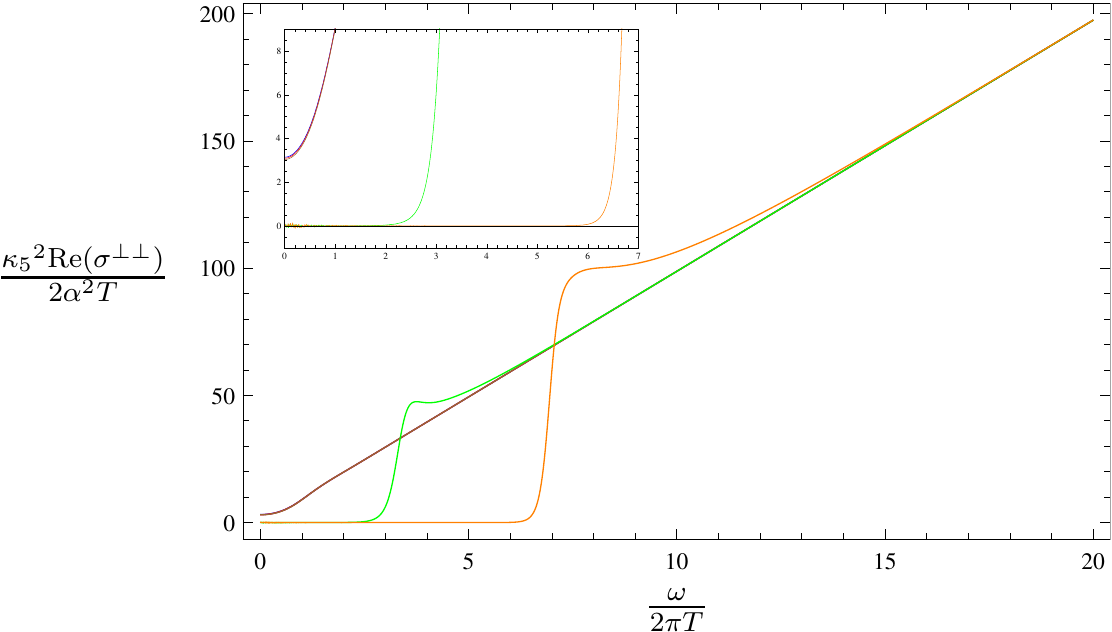}
\caption{Real part of the conductivity $\re(\sigma^{\perp\perp})$ over the frequency $\omega/(2\pi T)$ for $\alpha=0.032$. The color coding is as follows: blue $T=\infty$, red $T=1.34T_c$, brown $T=1.00T_c$, green $T=0.40T_c$, orange $T=0.19T_c$. Note that the three curves with the highest temperature, blue, red and brown, are nearly on top of each other. The agreement of the curves in the $\omega\to0$ limit is due to the small change in the strength of the Drude peak with temperature. Below $T_c$, the superfluid contribution to the delta peak at $\omega = 0$ is turned on and we obtain larger deviations from the $T=\infty$ curve, since the area below the curves has to be the same for all temperatures (sum rule). Furthermore the value for $\omega \to 0$ clearly asymptotes to $0$ with decreasing temperature.}
\label{fig:conda3y_0032}
\end{figure}

\begin{figure}[t]
\centering
\includegraphics[width=0.9\linewidth]{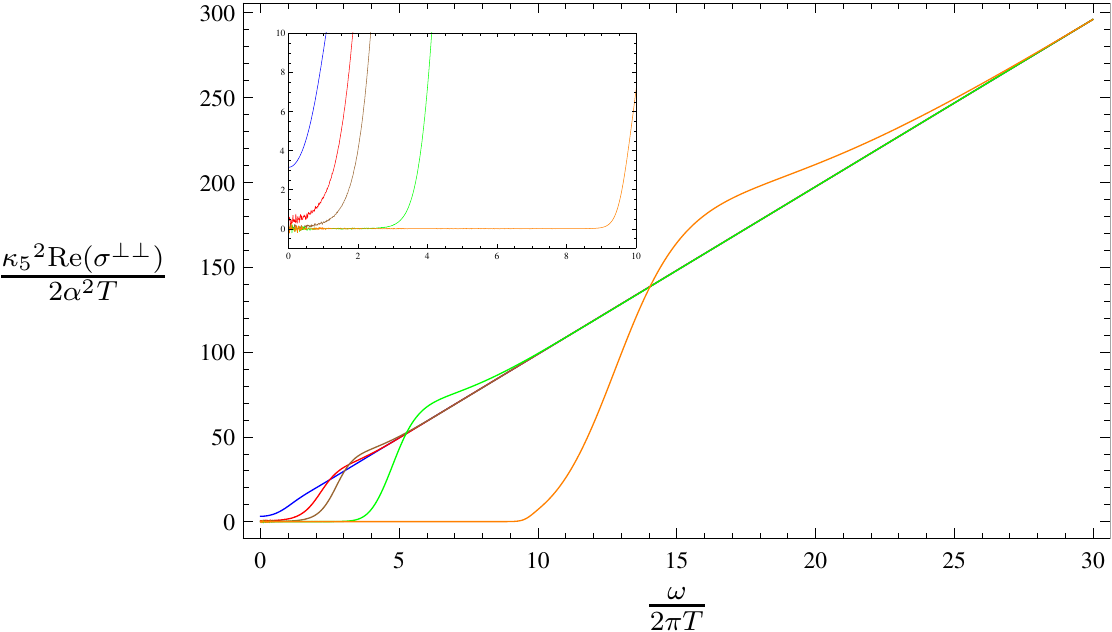}
\caption{Real part of the conductivity $\re(\sigma^{\perp\perp})$ over the frequency $\omega/(2\pi T)$ for $\alpha=0.316$. The color coding is as follows: blue $T=\infty$, red $T=1.00T_c$, brown $T=0.88T_c$, green $T=0.50T_c$, orange $T=0.19T_c$. In this plot we see that the Drude peak has already a much stronger dependence on the temperature than in the $\alpha=0.032$ case, since the blue and the red curve can be clearly distinguished. Below $T_c$ the contributions of the superfluid phase to the delta peak leads again to a tendency of the curve to vanish for frequencies in the gap region, since the area below the curves have to be the same for all temperatures (sum rule).}
\label{fig:conda3y_0316}
\end{figure}

Now we relate the results of \eqref{eq:hel1responseb1} to the thermoelectric effect on the field theory side. We begin with the well known connection between electric $\langle J_3^\perp\rangle=\langle J^\perp\rangle$ and thermal $\langle Q^\perp \rangle=\langle T^{t\perp}\rangle -\mu \langle J^\perp\rangle$ transport perpendicular to the condensate direction (see e.g. \cite{Hartnoll:2008kx,Hartnoll:2009sz,Herzog:2009xv} for the same calculation for holographic s-wave superfluids), i.e.
\begin{equation}
\label{eq:elthertrans}
\begin{pmatrix}\langle J^\perp\rangle\\ \langle Q^\perp \rangle\end{pmatrix}=\begin{pmatrix}\sigma^{\perp\perp} & T\alpha^{\perp\perp}\\ T\alpha^{\perp\perp} & T\bar{\kappa}^{\perp\perp} \end{pmatrix}\begin{pmatrix}E_\perp\\-(\nabla_\perp T)/T\end{pmatrix}\,,
\end{equation}
where the electric field $E_\perp$ and the thermal gradient $-\nabla_\perp T/T$ need to be related to the background values of the gauge field $\left(a^3_\perp\right)_0^b$ and the metric $\left(\Psi_t\right)^b_0$. This identification is done in \cite{Hartnoll:2009sz},
\begin{equation}
\label{eq:identtheroelectric}
\begin{split}
E_\perp&=\ii\omega\left(\left(a^3_\perp\right)_0^b+\mu \left(\Psi_t\right)^b_0\right)\,,\\
-\frac{\nabla_\perp T}{T}&=\ii\omega\left(\Psi_t\right)^b_0\,.
\end{split}
\end{equation}
Putting all together and comparing the relation of the electric and thermal transport to the corresponding equations for $\langle J^y\rangle$ and $\langle T^{ty}\rangle$ in (\ref{eq:hel1responseb1}), we can identify the transport matrix of (\ref{eq:elthertrans}),
   \begin{equation}
   \label{eq:trans3yty}
    \begin{split}
     \sigma^{\perp\perp}&=-\frac{\ii G^{\perp,\perp}_{3,3}}{\omega}=-\frac{\alpha^2r_h}{\kappa_5^2}\frac{\ii}{\tilde\omega}\left(\frac{2{\left(\tilde a_\perp^3\right)_1^b}}{{\left(\tilde a_\perp^3\right)_0^b}}-\frac{\tilde\omega^2}{2}\right)\,,\\
     T\alpha^{\perp\perp} &=-\frac{\ii}{\omega}\left({G^\perp_3}^{t\perp}-\mu G^{\perp,\perp}_{3,3}\right)= \frac{\ii}{\omega}\langle \calj^t_3\rangle-\mu\sigma^{\perp\perp},\\
     T\bar{\kappa}^{\perp\perp} &=-\frac{\ii}{\omega}\left(G^{t\perp,t\perp}+\mu^2G^{\perp,\perp}_{3,3}\right)=\frac{\ii}{\omega}\langle \calt_{tt} \rangle+\mu^2\sigma^{\perp\perp}.
    \end{split}
   \end{equation}
   These results are in agreement with \cite{Hartnoll:2009sz}. The coupling between thermal and electrical transport is well known in condensed matter physics, since the charge carriers (electron or holes) transport charge as well as heat. In this subset we do not observe any effect of the breaking of the rotational symmetry since all the fields are in the transverse direction to the condensate.
 
\begin{figure}[htb]
\centering
\includegraphics[width=0.9\linewidth]{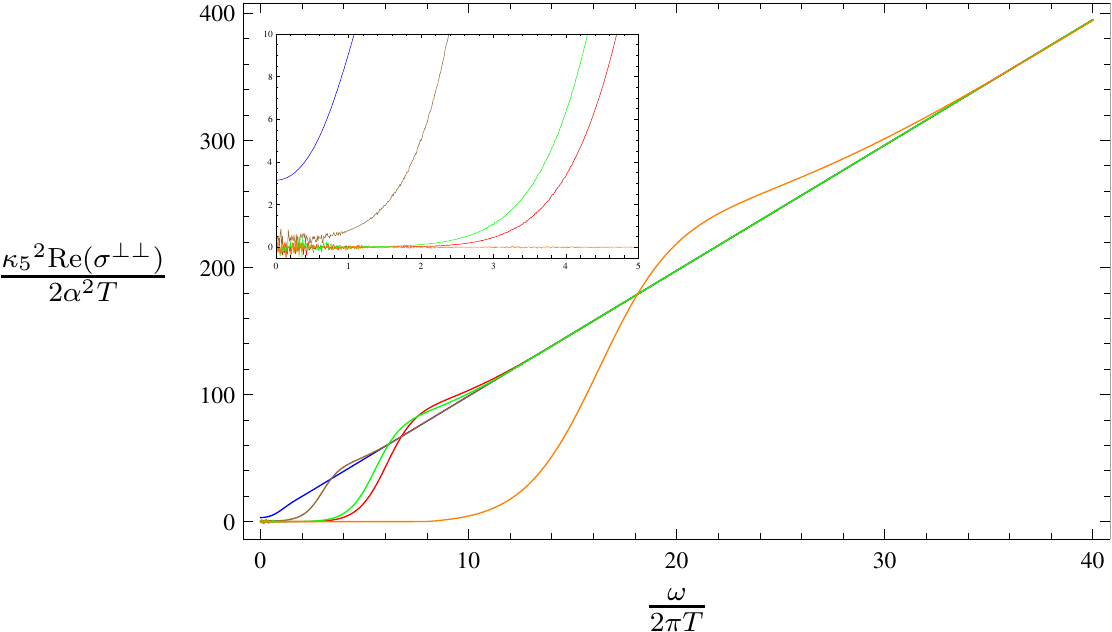}
\caption{Real part of the conductivity $\re(\sigma^{\perp\perp})$ over the frequency $\omega/(2\pi T)$ for $\alpha=0.447$. The color coding is as follows: blue $T=\infty$, brown $T=1.95T_c$, green $T=1.00T_c$, red $T=0.91T_c$, orange $T=0.34T_c$. Again we see the same tendency as before for the curve to vanish at $\omega\to0$ for decreasing temperatures. The strength of the Drude peak has a strong dependence on the temperatures, since the blue and the brown curve are quite far apart (both curves were computed for temperatures above $T_c$).}
\label{fig:conda3y_0447}
\end{figure}

   In figure \ref{fig:conda3y_0032}, \ref{fig:conda3y_0316} and \ref{fig:conda3y_0447} we plot our numerical results for $\re(\sigma^{\perp\perp})$ versus the frequency $\omega/(2\pi T)$ for different values of $\alpha$ as defined in \eqref{eq:alpha}, namely $\alpha=0.032<\alpha_c$, $\alpha=0.316\lesssim\alpha_c$ and $\alpha=0.447>\alpha_c$, respectively. For large frequencies, \ie $\omega\gg 2\pi T$, the conductivity asymptotically has a linear dependence on the frequency (e.g.~\cite{Myers:2007we}),
 \begin{equation}
 \label{eq:asymsigma3}
    \re(\sigma^{\perp\perp})\rightarrow \frac{\alpha^2}{{\kappa_5}^2}\pi \omega\qquad\text{for}\qquad \omega\gg 2\pi T\,.
\end{equation}
For small temperatures, \ie $T<T_c$, we see a gap opening up at small frequencies. The size of the gap increases as the temperature is decreased. This is the expected energy gap we know from superconductors. The gap ends at a frequency $\omega_g$ with a sharp increase of the conductivity. Beyond the gap the conductivity at small temperature, \ie $T<T_c$, is larger than the corresponding value at large temperature, \ie $T>T_c$ such that the small temperature conductivities approach the asymptotic behavior~\eqref{eq:asymsigma3} from above (cf. \cite{Gubser:2008wv,Ammon:2009fe,Ammon:2008fc}).

The value of $\re(\sigma^{\perp\perp})$ at $\omega=0$ approaches zero with decreasing temperature. Below $T_c$ the tendency for this part of the conductivity to vanish increases. Nevertheless, we still find finite values even below $T_c$, i.e. these values seem to be suppressed but not identically vanishing (c.f. \cite{Hartnoll:2008kx}). In \cite{Basu:2009vv} it is shown that in the limit $T\to0$ there is a hard gap, \ie the value for the conductivity becomes zero. Finally, we observe that an increase in $\alpha$ leads to a stronger suppression of the real part of the conductivity in the gap region.

Due to the sum rule for the conductivity, i.e. the frequency integral over the real part of the conductivity is constant for all temperatures, a delta peak has to form at zero frequency which contains the ``missing area'' of the gap region. The strength of the delta peak has two contributions: the first is proportional to the superfluid density $n_s$, $\re(\sigma^{\perp\perp})\sim \alpha^2/\kappa_5^2\; \pi n_s\delta(\omega)$ and appears only for temperatures below $T_c$. The second contribution is a consequence of translation invariance of our system, the Drude peak, and appears for all temperatures.

\begin{figure}[h]
\centering
\includegraphics[width=0.9\linewidth]{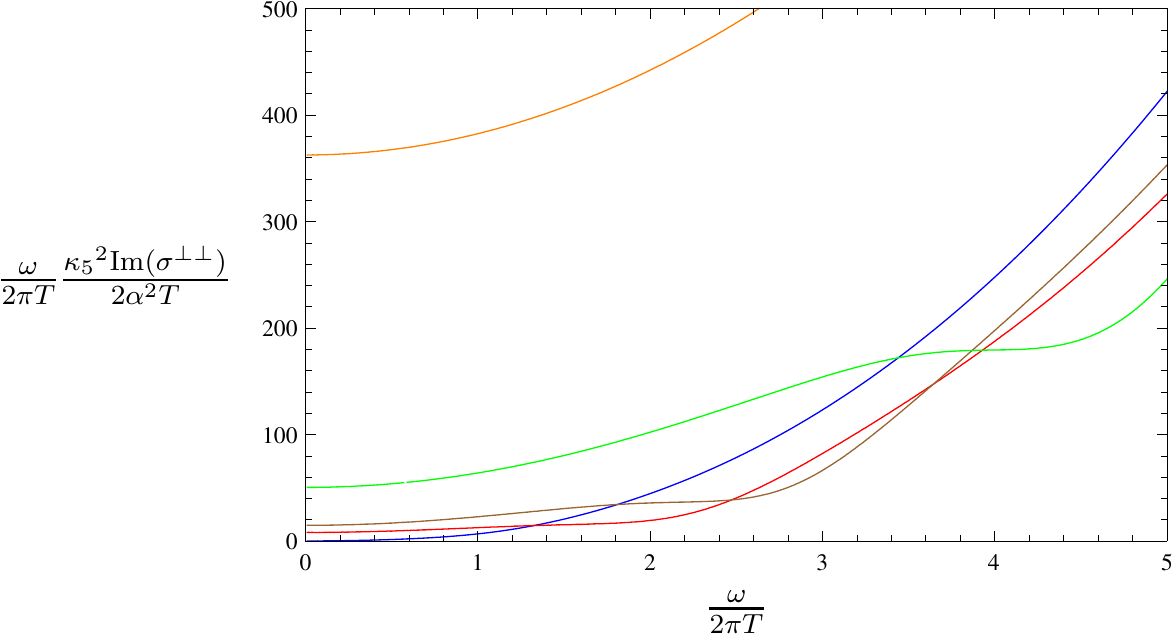}
\caption{Imaginary part of the conductivity $\omega \im(\sigma^{\perp\perp})$ over the frequency $\omega/(2\pi T)$ for $\alpha=0.316$. The color coding is as follows: blue $T=\infty$, red $T=1.00T_c$, brown $T=0.88T_c$, green $T=0.50T_c$, orange $T=0.19T_c$. The curves in this plot have a constant value for $\frac{\omega}{2\pi T}\to0$, which is determined by the $\delta$-peak in the real part of the conductivity by the Kramers-Kronig relation. The values for this constant are, in the same order as the temperatures above: 0, 8.0, 14.7, 50.3 and 362.4. Note that we already see a finite value for $T=1.00T_c$, this is due to the Drude peak. Below $T_c$ the values at $\omega=0$ are due to two contributions, first the Drude peak, as before, and second due to the superfluid density.}
\label{fig:conda3yim_0316}
\end{figure}

The delta peak is observed in the imaginary part of the conductivity by using the Kramers-Kronig relation (see \cite{Hartnoll:2008kx}),
\begin{equation}
\label{eq:imsigmaperpperp}
 \im(\sigma^{\perp\perp}) \simeq \frac{A_D (\alpha,T)}{\omega} + \frac{A_s(\alpha)}{\omega}\left(1-\frac{T}{T_c}\right),
\end{equation}
for $T \lesssim T_c$, with $A_s(\alpha)\left(1-\frac{T}{T_c}\right) \propto n_s$ and $A_D$ parametrizing the contribution from the Drude peak. In figure~\ref{fig:conda3yim_0316} we present the imaginary part of the conductivity $\omega \im(\sigma^{\perp\perp})$ versus the frequency $\omega/(2\pi T)$ for $\alpha=0.316$ and different temperatures. We see that $\omega \im(\sigma^{\perp\perp})$ takes finite values for $\omega\to0$ and $T<\infty$. The finite values above $T_c$ are due to the Drude peak, i.e. the $A_D$ part of (\ref{eq:imsigmaperpperp}). Below $T_c$ we see a further contribution from the superfluid density. By analyzing the temperature dependence of $\lim_{\omega\to0} \omega \im(\sigma^{\perp\perp})${\bf ,} we get a smooth curve, which is, however, not differentiable at $T_c$ for $\alpha \leq \alpha_c$, i.e. it behaves as equation (\ref{eq:imsigmaperpperp}) anticipated. However, for $\alpha > \alpha_c$ we see a jump at $T_c$ as consequence of the jump in the condensate. Furthermore $A_D$ and $A_s$ have a non trivial dependence on $\alpha$. Finally, as expected, there is an increase in the superfluid density with decreasing temperature.

\subsection{Non-Universal Shear Viscosity and Flexoelectric Effect}
\label{sec:Non-Universal-Shear-Viscosity}
\begin{figure}[t]
\centering
\includegraphics[width=0.9\linewidth]{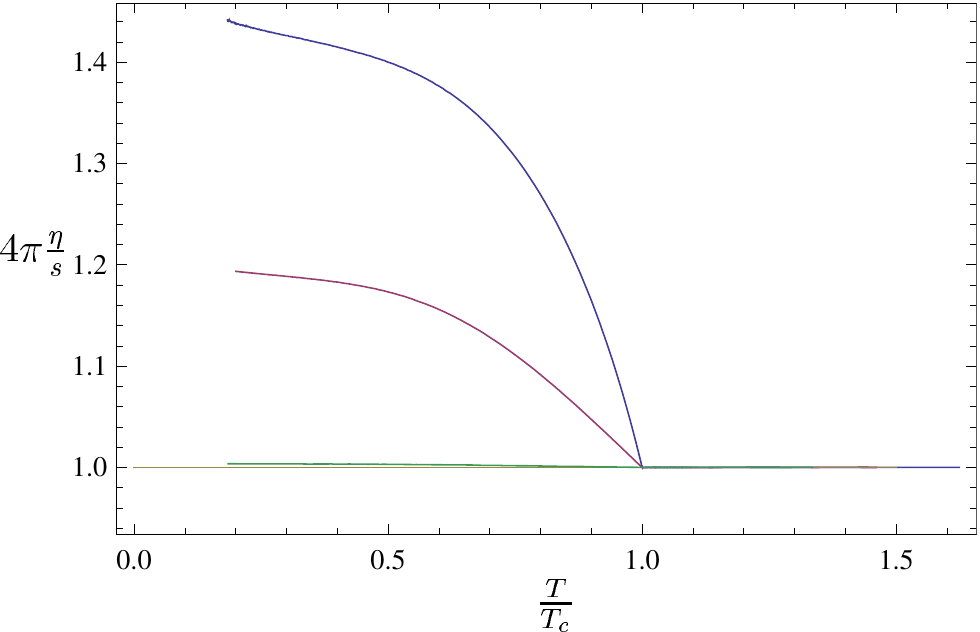}
\caption{Ratio of shear viscosities $\eta_{yz}$ and $\eta_{x\perp}$ to
  entropy density $s$ over the reduced temperature $T/T_c$ for
  different values of the ratio of the gravitational coupling constant
  to the Yang-Mills coupling constant $\alpha$. The color coding is as
  follows: In yellow, $\eta_{yz}/s$ for all values of 
$\alpha$; while the curve for $\eta_{x\perp}/s$ is plotted in green for $\alpha=0.032$, red for $\alpha=0.224$ and blue for $\alpha=0.316$. The shear viscosities coincide and are universal in the normal phase $T\ge T_c$. However in the superfluid phase $T<T_c$, the shear viscosity $\eta_{yz}$ has the usual universal behavior while the shear viscosity $\eta_{x\perp}$ is non-universal.}
\label{fig:viscosity}
\end{figure}

Now let us study the remaining three components of the helicity one modes, $\langle J^{\perp}_1\rangle$, $\langle J^{\perp}_2\rangle$ and $\langle T^{x\perp}\rangle$ as given by
\eqref{eq:hel1responseb2}. We first focus on $\langle T^{x\perp}\rangle$, which for $\left(a_y^1\right)_0^b,\left(a_y^2\right)_0^b=0$ can be translated into the following dual field theory behavior
\begin{equation}
\langle T^{x\perp}\rangle= -\langle \calt_{xx}\rangle h_{x\perp} -i\omega \eta_{x\perp} h_{x\perp}\,,
\end{equation}
where $\eta_{x\perp}$ is the second shear viscosity which is present in a transversal isotropic fluid (see appendix~\ref{sec:General-Remarks-on}) and with $\langle \calt_{xx} \rangle$ as defined in \eqref{eq:cftstressenergytensor}. Here we see again that we can apply the Kubo formula to determine the shear viscosity $\eta_{x\perp}$,
\begin{equation}
\eta_{x\perp} = -\lim_{\omega\to0}\frac{1}{\omega}\im\left(G^{x\perp,x\perp}\right)\,. \label{eq:corretaxy}
\end{equation}
As described in \cite{Erdmenger:2010xm}, this shear viscosity has a non-trivial temperature dependence even in the large $N$ and large $\lambda$ limit and is therefore not universal. In fig.~\ref{fig:viscosity} we compare our numerical results for the
ratio of the shear viscosity $\eta_{x\perp}$ to the entropy density $s$
with the universal behavior of the shear viscosity $\eta_{yz}$ for
different values of $\alpha$. We see that in the normal
phase $T\ge T_c$, the two shear viscosities coincide as required in an
isotropic fluid. In addition, the ratio of shear viscosity to entropy
density is universal. In the superfluid phase $T<T_c$, the two shear
viscosities deviate from each other and $\eta_{x\perp}$ is
non-universal. However it is exciting that $\eta_{x\perp}/s \geq 1/4\pi$, 
such that the KSS bound on the ratio
of shear viscosity to entropy density \cite{Kovtun:2004de} is still
valid.

\begin{figure}[t]
\centering
\includegraphics[width=0.9\linewidth]{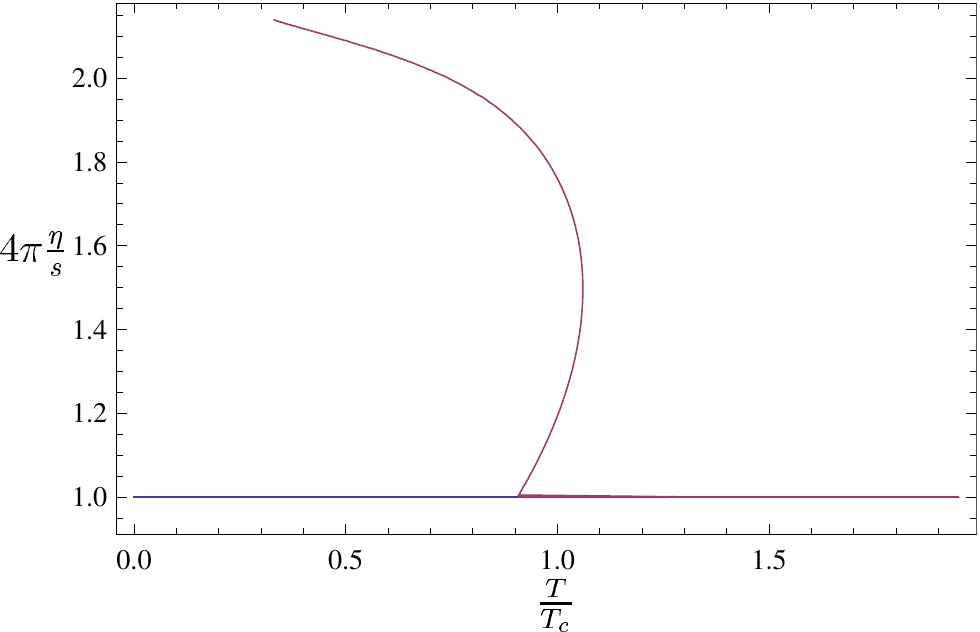}
\caption{Ratio of shear viscosities $\eta_{yz}$ (blue) and $\eta_{x\perp}$ (red) to entropy density $s$ over the reduced temperature $T/T_c$ for $\alpha=0.447$, which is larger than the critical value where the phase transition becomes first order:  The shear viscosities coincide in the normal phase $T\ge T_c$ and are universal. In the superfluid phase $\eta_{x\perp}$ is non-universal. Close to the phase transition, it is multivalued as expected  for a first order phase transition.}
\label{fig:viscosityfirstorder}
\end{figure}

The difference between the two viscosities in the superfluid
phase is controlled by $\alpha$ as defined in \eqref{eq:alpha}. In the probe limit where $\alpha=0$,
the shear viscosities also coincide in the superfluid phase. By
increasing the back-reaction of the gauge fields, \ie rising $\alpha$,
the deviation between the shear viscosities becomes larger in the
superfluid phase as shown in fig.~\ref{fig:viscosity}. If $\alpha$ is
larger than the critical value $\alpha_c=0.365$ found in \cite{Ammon:2009xh} (see fig.~\ref{fig:phasediag}) where the
phase transition to the superfluid phase becomes first order, the shear viscosities are also multivalued close to the phase transition as seen in fig.~\ref{fig:viscosityfirstorder}.  Since there is a maximal $\alpha$ denoted by $\alpha_\text{max}=0.628$ for which the superfluid phase exists (see fig.~\ref{fig:phasediag}), we expect that the deviation of the shear viscosity $\eta_{x\perp}$ from its universal value is maximal for this $\alpha_\text{max}$. Unfortunately numerical calculations for large values of $\alpha$ are very challenging such that we cannot present satisfying numerical data for this region. It is interesting that also the deviations due to $\lambda$ and $N_c$ corrections are bounded. In this case the bound is determined by causality \cite{Buchel:2009tt}.

As described in our letter \cite{Erdmenger:2010xm}, we also have found numerically that for $\alpha<\alpha_c$
\begin{equation}
1 - 4 \pi \frac {\eta_{x \perp}}{s} \propto \left(1 - \frac{T}{T_c} \right)^\beta \quad {\mathrm{ with}} \quad 
\beta = 1.00 \pm 3 \% \, .
\end{equation}
Interestingly, the value of $\beta$ appears to be independent of $\alpha$.
This result has recently been confirmed by an analytic calculation in \cite{Basu:2011tt}.

The non-universality of the shear viscosity can be understood in the following way. For the $\eta_{x \perp}$ component, the equation of motion \eqref{eq:eompwavefluchxy} in the $\omega\to0$ limit includes also non-vanishing source terms besides the derivative of the conjugate momentum $\Pi_x$ of $\Psi_x$, i.e. $\partial_r \Pi_x = \mathrm{source}$. This is in contrast to equation of motion which leads to the
$\eta_{yz}$ component. Note that the source term depends on the condensate $w$ and the fluctuation $a^1_\perp$ and vanishes if the condensate $w$ vanishes. Hence, as we confirm numerically in fig.~\ref{fig:viscosity}, when the condensate is absent (i.e. for the $T>T_c$ case) we obtain again the universal result, since in this case the same proof as described above for the helicity two mode applies.

For $h_{x\perp}=0$, we obtain flavor charge transport, \ie a flavor field $a^{1,2}_\perp$ generates a flavor current $\langle J^\perp_{1,2}\rangle$. In the unbroken phase it is useful to combine the fields $a_\perp^{1,2}$ in the way $a^\pm_\perp=a^1_\perp\pm \ii a^2_\perp$ since they transform in the fundamental representation of the $U(1)_3$ symmetry and are complex conjugate to each other. To make contact to the unbroken phase, we also use this definition in the broken phase.

We also use the definition for the currents $\langle J_\pm\rangle=1/2\left(\langle J_1\rangle\pm\ii \langle J_2\rangle\right)$, such that the full transport matrix becomes
\begin{equation}
\label{eq:hel1flexshear}
\begin{pmatrix}\langle J^{\perp}_+\rangle\\[1ex] \langle J^{\perp}_-\rangle\\[1ex] \langle T^{x\perp}\rangle \end{pmatrix}=
\begin{pmatrix}
      G_{+,+}^{\perp,\perp} & G_{+,-}^{\perp,\perp} & {G_{+}^{\perp}}^{x\perp}\\[1ex]
      G_{-,+}^{\perp,\perp} & G_{-,-}^{\perp,\perp} & {G_{-}^{\perp}}^{x\perp}\\[1ex]
      {G^{x\perp}}_{+}^{\perp}\quad & {G^{x\perp}}_{-}^{\perp}\quad & -\langle T_{xx}\rangle-\ii\omega\eta_{x\perp}
\end{pmatrix}
\begin{pmatrix}
      a^+_\perp\\[1ex]
      a^-_\perp\\[1ex]
      h_{x\perp}
\end{pmatrix}\,,
\end{equation}
where the flavor conductivities are given by
\begin{equation}
\label{eq:condflavor}
\begin{split}
     G_{\pm,\pm}^{\perp,\perp} (\omega)&=\frac{1}{4}\left[G^{\perp,\perp}_{1,1}(\omega)+G^{\perp,\perp}_{2,2}(\omega)\mp\ii\left(G^{\perp,\perp}_{1,2}(\omega)-G^{\perp,\perp}_{2,1}(\omega)\right)\right]\\
     &=\frac{\alpha^2r_h^2}{2\kappa_5^2}\left(\frac{{\left(\tilde a_\perp^1\right)_1^b}(\omega)}{{\left(\tilde a_\perp^1\right)_0^b}(\omega)}+\frac{{\left(\tilde a_\perp^2\right)_1^b}(\omega)}{{\left(\tilde a_\perp^2\right)_0^b}(\omega)}-\frac{\left(\tilde\mu\mp\tilde\omega\right)^2}{2}\mp\ii\left(\frac{{\left(\tilde a_y^1\right)_1^b}(\omega)}{{\left(\tilde a_y^2\right)_0^b}(\omega)}-\frac{{\left(\tilde a_y^2\right)_1^b}(\omega)}{{\left(\tilde a_y^1\right)_0^b}(\omega)}\right)\right)\,,\\
     G_{\pm,\mp}^{\perp,\perp}(\omega) &=\frac{1}{4}\left[G^{\perp,\perp}_{1,1}(\omega)-G^{\perp,\perp}_{2,2}(\omega)\pm\ii\left(G^{\perp,\perp}_{1,2}(\omega)+G^{\perp,\perp}_{2,1}(\omega)\right)\right]\\
     &=\frac{\alpha^2r_h^2}{2\kappa_5^2}\left(\frac{{\left(\tilde a_\perp^1\right)_1^b}(\omega)}{{\left(\tilde a_\perp^1\right)_0^b}(\omega)}-\frac{{\left(\tilde a_\perp^2\right)_1^b}(\omega)}{{\left(\tilde a_\perp^2\right)_0^b}(\omega)}\pm\ii\left(\frac{{\left(\tilde a_y^1\right)_1^b}(\omega)}{{\left(\tilde a_y^2\right)_0^b}(\omega)}+\frac{{\left(\tilde a_y^2\right)_1^b}(\omega)}{{\left(\tilde a_y^1\right)_0^b}(\omega)}\right)\right)\,,\\
     {G^{x\perp}}_\pm^\perp(\omega)&=\frac{1}{2}\left[{G^{x\perp}}_{1}^{\perp}(\omega)\mp\ii{G^{x\perp}}_{2}^{\perp}(\omega)\right]=-\frac{\langle \calj_1^x\rangle}{4} + \frac{r_h^3}{\kappa_5^2}\left(\frac{{\big(\Psi_x\big)_2^b}(\omega)}{{\left(\tilde a_y^1\right)_0^b}(\omega)}\mp\ii \frac{{\big(\Psi_x\big)_2^b}(\omega)}{{\left(\tilde a_y^2\right)_0^b}(\omega)}\right)\,,\\
     {G_\pm^\perp}^{x\perp}(\omega)&=\frac{1}{2}\left[{G_{1}^{\perp}}^{x\perp}(\omega)\pm\ii{G_{2}^{\perp}}^{x\perp}(\omega)\right]=-\frac{\langle \calj_1^x\rangle}{4} + \frac{\alpha^2r_h^3}{\kappa_5^2}\left(\frac{{\left(\tilde a_y^1\right)_1^b}(\omega)}{{\big(\Psi_x\big)_0^b}(\omega)}\pm\ii \frac{{\left(\tilde a_y^2\right)_1^b}(\omega)}{{\big(\Psi_x\big)_0^b}(\omega)}\right)\,.
\end{split}
\end{equation}
First note that for $\mu=0$ where the $SU(2)$ symmetry is restored, \ie $a^1\equiv a^2$, the Green's function is diagonal and $G_{+,+}^{\perp,\perp}=G_{-,-}^{\perp,\perp}=G_{3,3}^{\perp,\perp}$. In the unbroken phase, $G_{+,-}^{\perp,\perp}=G_{-,+}^{\perp,\perp}\equiv 0$ is still valid for $\mu\not=0$, since the $a^\pm$ do not couple to each other, while $G_{+,+}^{\perp,\perp}\not=G_{-,-}^{\perp,\perp}$ for $\mu\not=0$. In the unbroken as well as in the broken phase we find
\begin{equation}
\begin{split}
\label{eq:symsigmapm}
  &G_{-,-}^{\perp,\perp}(\omega)=G_{+,+}^{\perp,\perp}(-\omega)^*\,,\quad\,\,\,\qquad G_{+,-}^{\perp,\perp}(\omega)=G_{-,+}^{\perp,\perp}(-\omega)^*\,,\\
  &{G_+^\perp}^{x\perp}(\omega)={G_-^\perp}^{x\perp}(-\omega)^*\quad \text{and} \quad {G^{x\perp}}_+^\perp(\omega)={G^{x\perp}}_-^\perp(-\omega)^*\,,
\end{split}
\end{equation}
as expected since $a^1(\omega)=\left(a^1(-\omega)\right)^*$, $a^2(\omega)=\left(a^2(-\omega)\right)^*$ and $\Psi_x(\omega)=\left(\Psi_x(-\omega)\right)^*$.

\begin{figure}
  \centering
  \subfigure[]{\label{fig:remm0316}\includegraphics[width=0.45\textwidth]{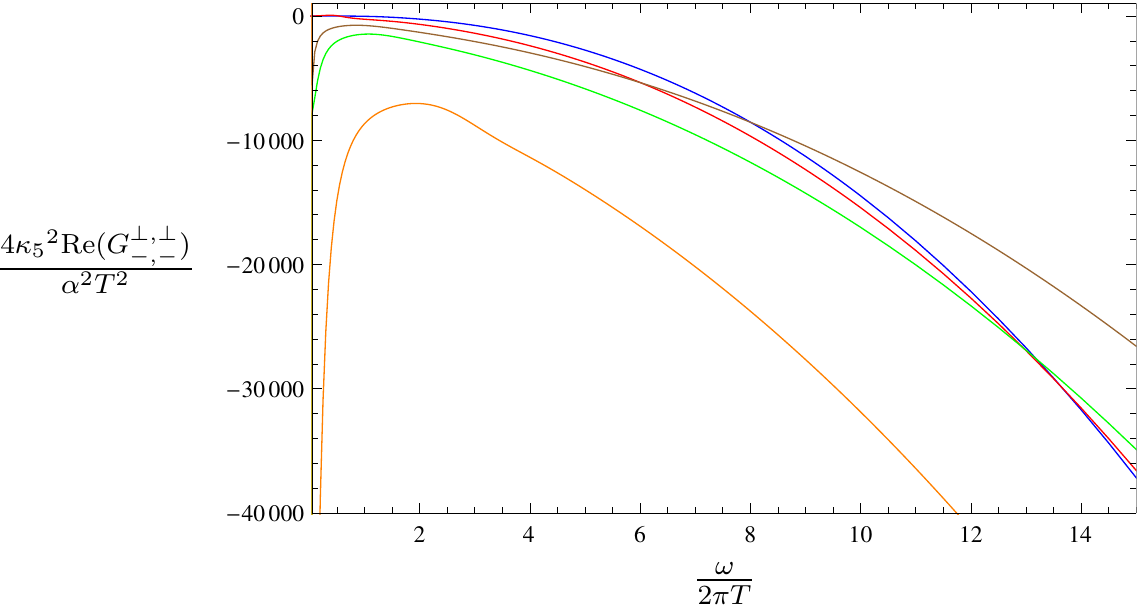}}
  \hfill               
  \subfigure[]{\label{fig:repp0316}\includegraphics[width=0.45\textwidth]{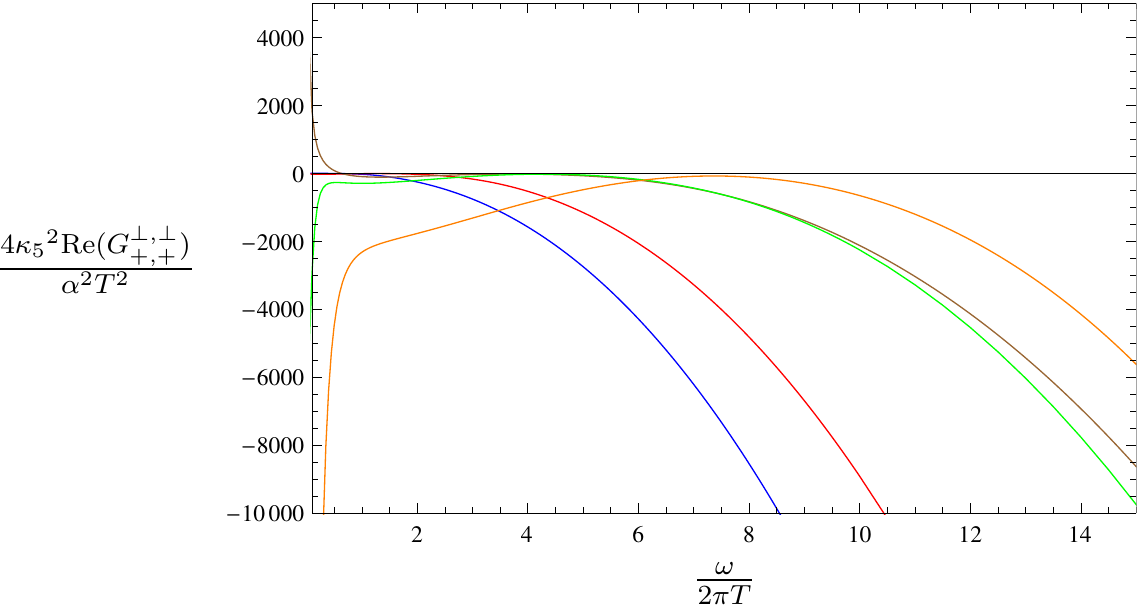}}
  \hfill               
  \subfigure[]{\label{fig:immm0316}\includegraphics[width=0.45\textwidth]{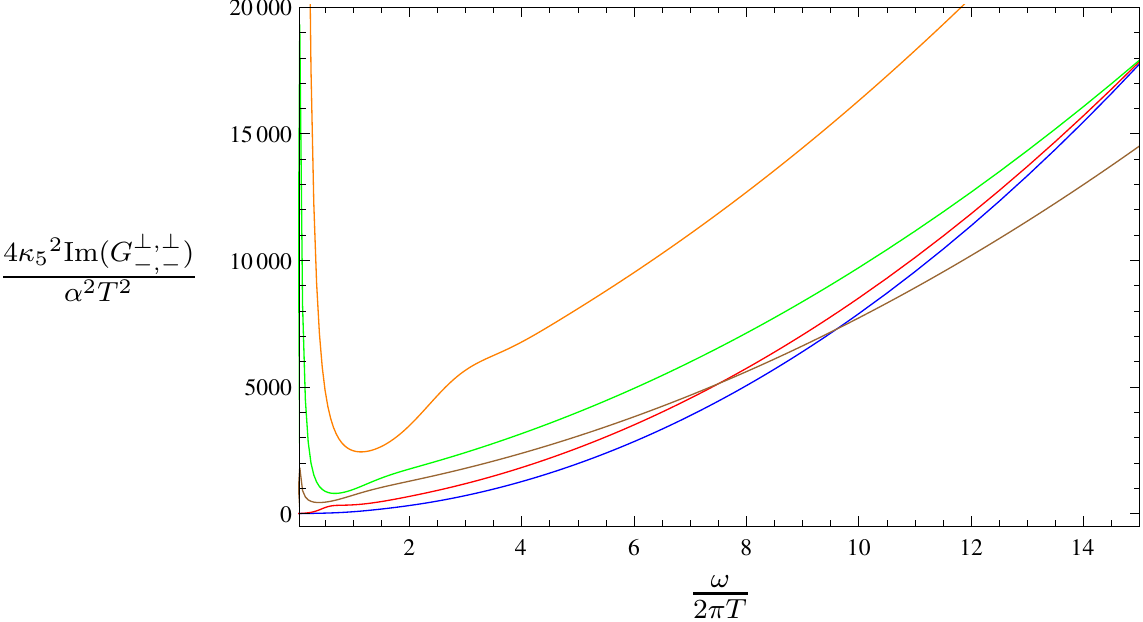}}
  \hfill               
  \subfigure[]{\label{fig:impp0316}\includegraphics[width=0.45\textwidth]{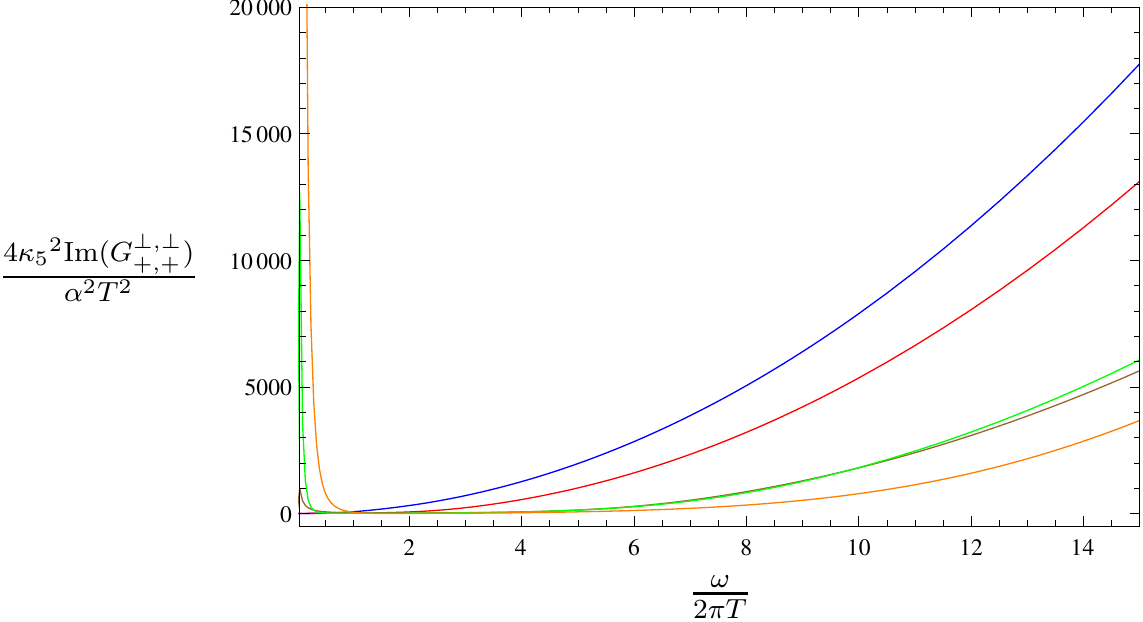}}
  \caption{These plots show the real and imaginary part of the correlators $G^{\perp,\perp}_{\pm,\pm}$ versus the reduced frequency $\omega/(2\pi T)$ for $\alpha=0.316$ at different temperatures: $T=\infty$ blue line, $T=3.02T_c$ red line, $T=1.00T_c$ brown line, $T=0.88T_c$ green line and $T=0.50T_c$ orange line.}
  \label{fig:corrppmm}
\end{figure}

\begin{figure}
  \centering
  \subfigure[]{\label{fig:repm0316}\includegraphics[width=0.45\textwidth]{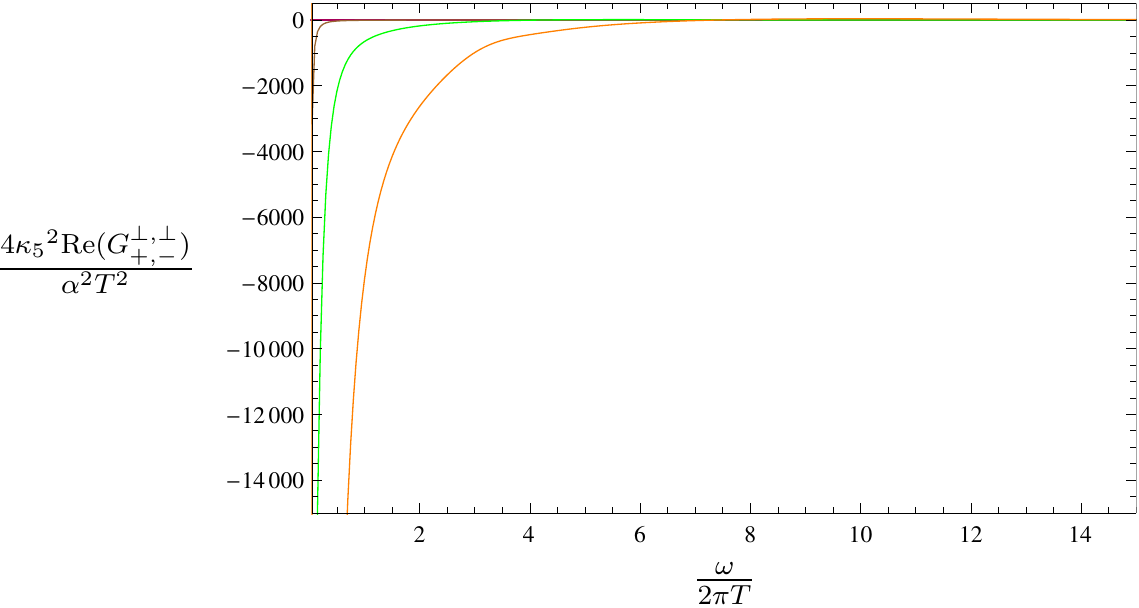}}                
  \hfill
  \subfigure[]{\label{fig:impm0316}\includegraphics[width=0.45\textwidth]{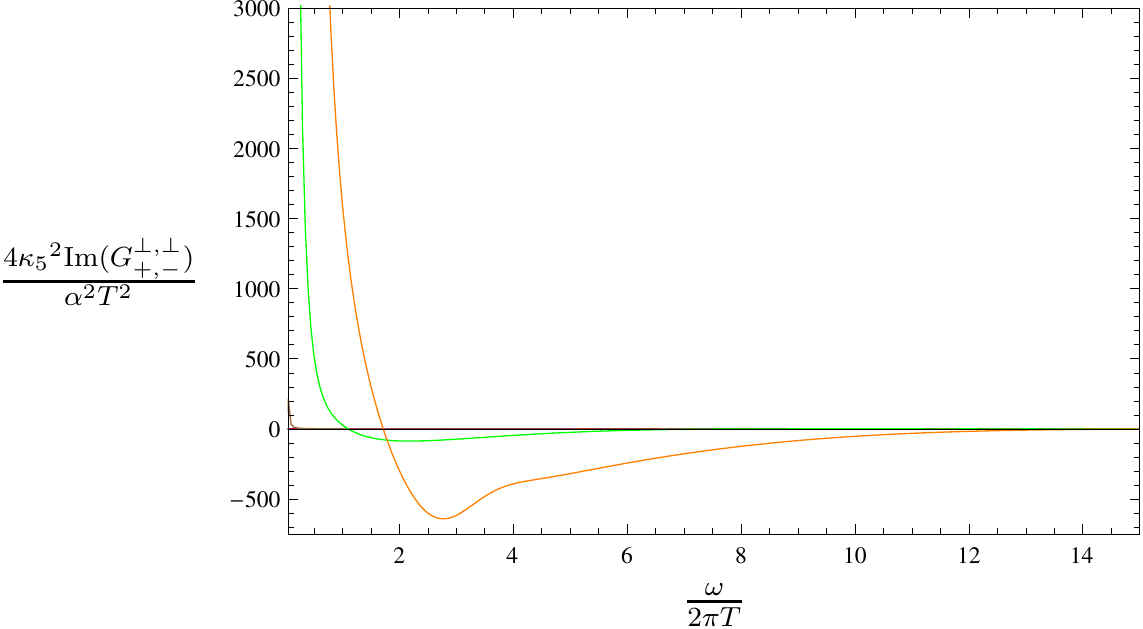}}
  \caption{These plots show the real and imaginary part of the correlator $G^{\perp,\perp}_{+,-}$ versus the reduced frequency $\omega/(2\pi T)$ for $\alpha=0.316$ at different temperatures: $T=\infty$ blue line, $T=3.02T_c$ red line, $T=1.00T_c$ brown line, $T=0.88T_c$ green line and $T=0.50T_c$ orange line. The curves for the temperatures above $T_c$ are exactly zero for all frequencies.}
  \label{fig:corrpmmp}
\end{figure}

\begin{figure}
  \centering
  \subfigure[]{\label{fig:repxy0316}\includegraphics[width=0.45\textwidth]{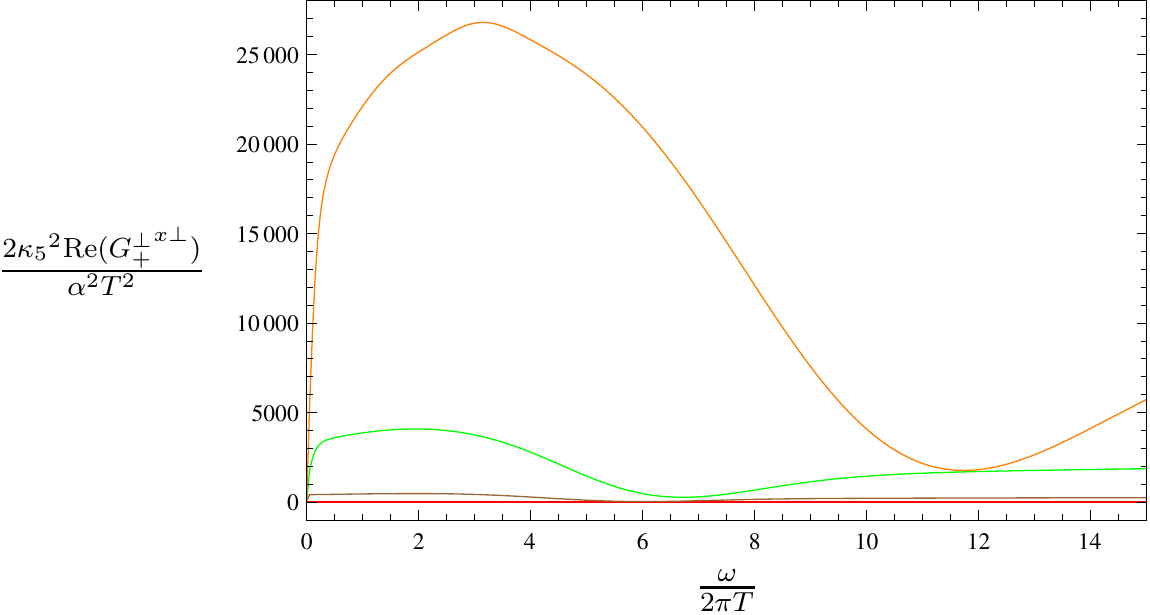}}                
  \hfill
  \subfigure[]{\label{fig:impxy0316}\includegraphics[width=0.45\textwidth]{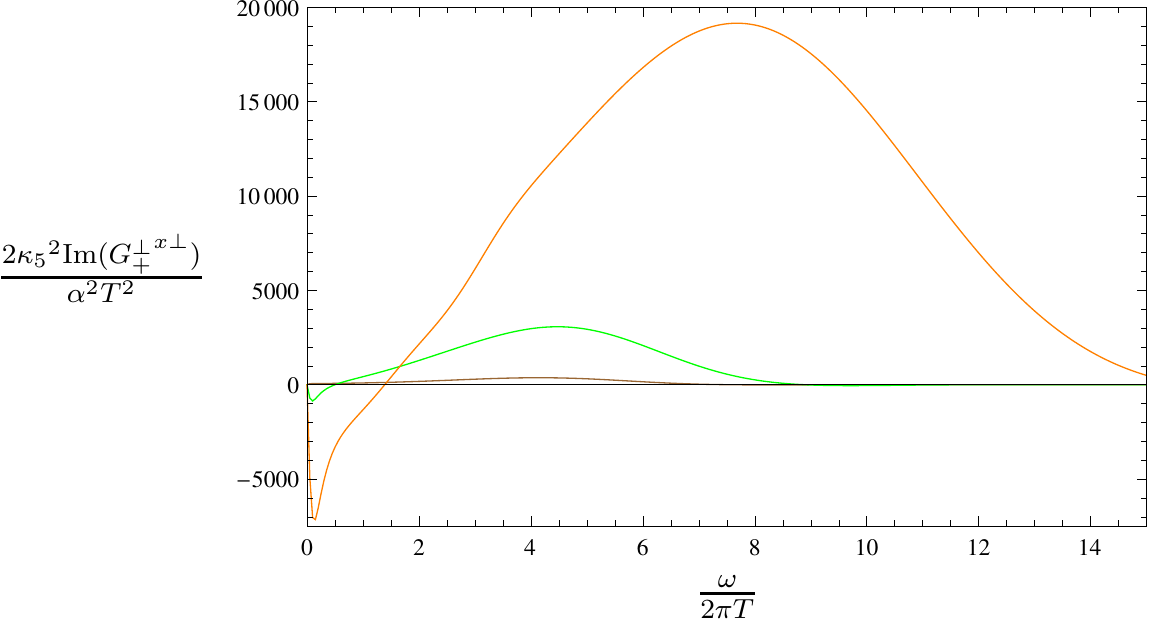}}                
  \hfill
  \subfigure[]{\label{fig:remxy0316}\includegraphics[width=0.45\textwidth]{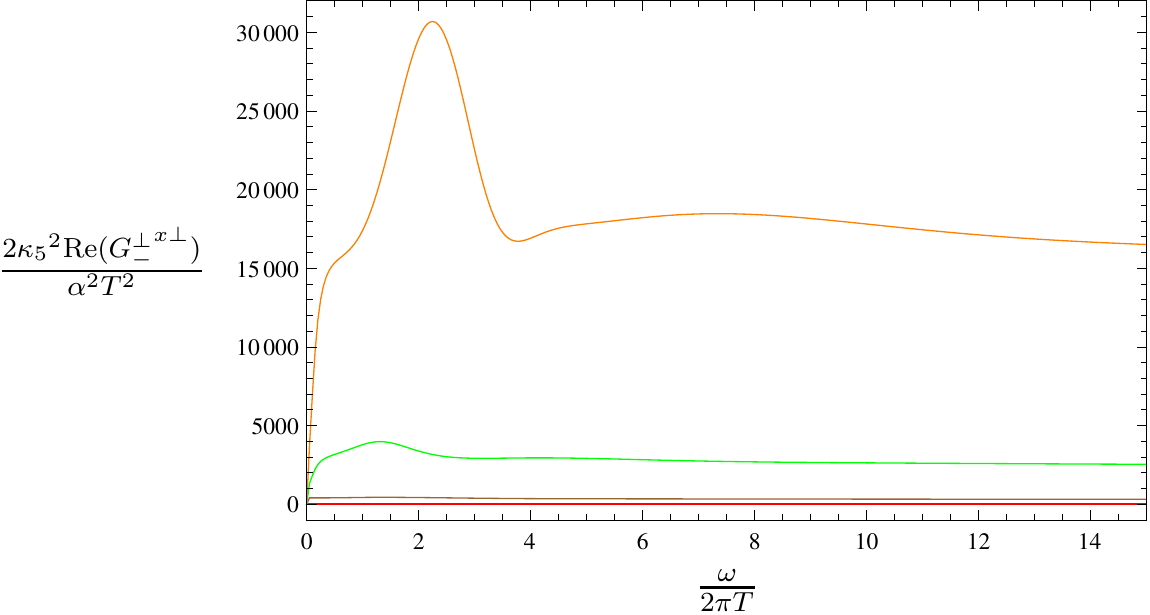}}                
  \hfill
  \subfigure[]{\label{fig:immxy0316}\includegraphics[width=0.45\textwidth]{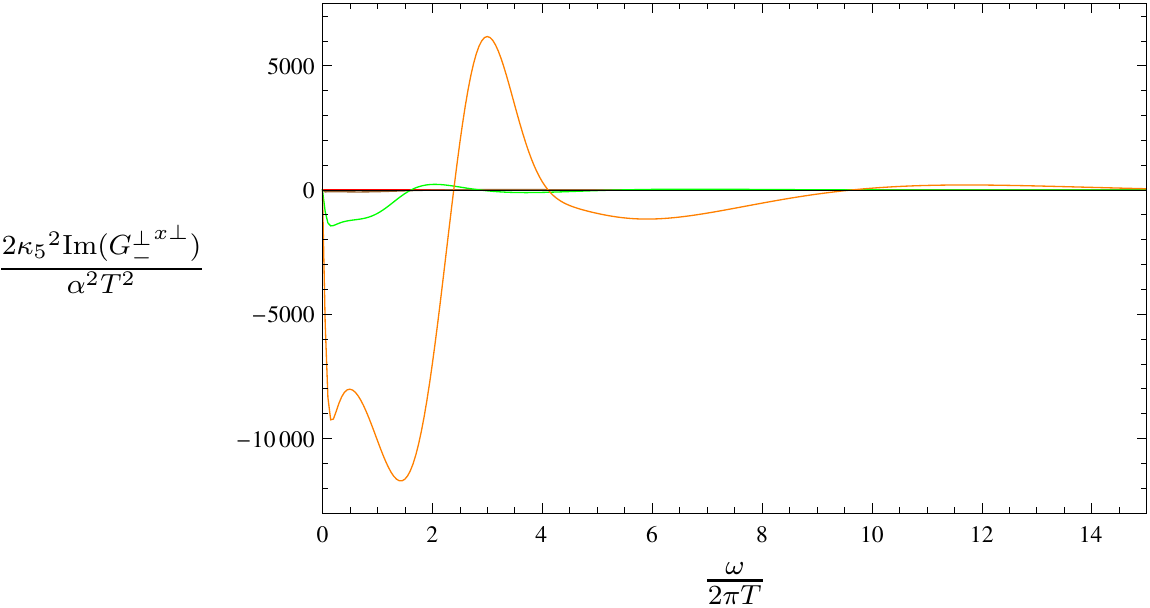}}
  \caption{These plots show the real and imaginary part of the correlators ${G^\perp_\pm}^{x\perp}$ versus the reduced frequency $\omega/(2\pi T)$ for $\alpha=0.316$ at different temperatures: $T=\infty$ blue line, $T=3.02T_c$ red line, $T=1.00T_c$ brown line, $T=0.88T_c$ green line and $T=0.50T_c$ orange line. The curves for the temperatures above $T_c$ are exactly zero for all frequencies.}
  \label{fig:corrpmxy}
\end{figure}

In figure \ref{fig:corrppmm} we plot the real and imaginary parts of $G_{\pm,\pm}^{\perp,\perp}(\omega)$.  We see in fig.~\ref{fig:immm0316} and \ref{fig:impp0316}, showing $\im(G^{\perp,\perp}_{\pm,\pm})$, that for temperatures $T>T_c$, the quasinormal modes tend towards the origin (in these plots we see their projection on the real axis) (see \eg \cite{Gubser:2008wv,Erdmenger:2008yj}). For $T\leq T_c$ we see a pole at the origin which is due to the massless Nambu-Goldstone modes. These Nambu-Goldstone modes are related to rotations of the director $\langle\calj_x^1\rangle$ in real space which are generated by the fluctuations $a_\perp^1$ \footnote{The other Nambu-Goldstone mode is related to the change of the phase of the condensate and correspond to the fluctuation $a_x^2$ which shows up in the helicity zero sector.}. Furthermore, as expected for large frequencies, the Green's function grows proportional to $\omega^2$ in the $\pm\pm$ components as for the correlator $G^{\perp,\perp}_{3,3}$. In figure \ref{fig:corrppmm} the correlators for the different temperatures do not seem to have the same asymptotic behavior. However, in the present case we have contribution from terms such as $\omega\mu$, i.e. of first order in $\omega$, which are not existent in the $G^{\perp,\perp}_{3,3}$ component. Hence to see that all correlators have the same limit, larger values of $\omega$ have to be considered. Even not present in our figures we verified numerically that the asymptotics of the correlators at different temperatures agree. Furthermore, in this context, we check that the fluctuations  $a^\pm_\perp$, which are unstable in the normal phase, are stabilized in the broken phase for $T<T_c$, i.e. the quasinormal modes of these perturbations stay in the lower half plane. Thus, the preferred direction induced by the current $\langle \calj^x_1\rangle$ is stable\footnote{Note that the Coleman-Mermin-Wagner-Hohenberg theorem does not apply here since the field theory is in 3+1 dimensions.}. A more detailed study of this sector in the probe approximation is in preparation \cite{Barisch:2011TA}.

In figure \ref{fig:corrpmmp} we plot $G_{+,-}^{\perp,\perp}(\omega)$. We see that this correlator vanishes since the $a^\pm$ do not couple in the unbroken phase. Furthermore, below $T_c$, a pole at $\omega=0$ due to the Nambu-Goldstone mode appears. We do not show $G_{-,+}^{\perp,\perp}(\omega)$ since $G_{+,-}^{\perp,\perp}$ and $G_{-,+}^{\perp,\perp}$ look alike.
Nevertheless, there is a difference between them in the broken phase. The difference arises from the contributions to the correlators due to the mixed terms, $G_{1,2}^{\perp,\perp}$ and $G_{2,1}^{\perp,\perp}$, in the corresponding equation in \eqref{eq:condflavor}. However, these are suppressed in relation  to $G_{1,1}^{\perp,\perp}$, which contains the Goldstone mode in the broken phase.

Finally, in figure \ref{fig:corrpmxy} we show ${G_\pm^\perp}^{x\perp}(\omega)$. Note that the contribution of $\langle \calj^x_1 \rangle$ to the real parts of the correlators is not included in the corresponding plots since it just shifts the curves by a constant. Furthermore, we have that $\im({G^\perp_\pm}^{x\perp})=\im({G^{x\perp}}_\pm^\perp)$ and $\re({G^\perp_\pm}^{x\perp})-\mathrm{const}=\re({G^{x\perp}}_\pm^\perp)$, i.e. there is a constant offset between the real parts of these correlators. We expect that this constant offset is generated by the term ${\big(\Psi_x\big)_2^b}(\omega)/{\left(\tilde a_y^1\right)_0^b}(\omega)$ which may be constant in the limit $\omega\to0$ since $a_y^1$ has a normal mode at $\omega=0$ and the subleading term of $\Psi_x$ is probably not sourced.  A analytic calculation similar to \cite{Basu:2011tt} may confirm this claim. In the unbroken phase these correlators vanish since the differential equations of the corresponding fields decouple. In the broken phase the correlators present a rich structure, which we cannot fully address at present. However, it seems that the coupling between the $a^\pm_\perp$ flavor fields and the strain $h_{x\perp}$ generates new quasiparticles which appear as bumps in the curves.

In addition to the flavor conductivity and the shear viscosity we obtain a coupling between the stress $ \langle T^{x\perp}\rangle$ and the flavor fields $a^\pm_\perp$ as well as the currents $\langle J^{\perp}_\pm\rangle$ and the strain $h_{x\perp}$ described in~\eqref{eq:hel1flexshear}. This coupling introduces an effect which is called flexoelectric effect in nematic crystals \cite{Gennes:1974lc} and only appears in fluids with broken rotational symmetry. We have a current $\langle \calj^{x}_1\rangle$ in a favored direction in the background which interacts with the flavor fields $a^\pm_\perp$. This interaction induces a force on the current which pushes the current in its perpendicular direction generating the stress $ \langle T^{x\perp}\rangle$. In the similar way, a strain $h_{x\perp}$ introduces an inhomogeneity in the current $\langle \calj^{x}_1\rangle$ resulting in a flavor field $a^\pm_\perp$ which generates the currents $\langle J^\perp_\pm\rangle$.


\section{Conclusion}
\label{sec:Conclusion} 
In this paper we have studied transport phenomena in holographic p-wave superfluids constructed in the $SU(2)$ Einstein-Yang-Mills theory. We classify the perturbations about equilibrium according to their transformation properties under the symmetry group. At zero momentum, there is an $SO(2)$ symmetry left which allows us to divide the perturbations into different helicity sectors: helicity two, one and zero states. While the helicity two state is trivial and leads to the universal ratio of shear viscosity to entropy density, the helicity one states are non-trivial. Due to a $\mathbb{Z}_2$ parity, this sector splits into two blocks. In the first block we find the thermoelectric effect transversal to the direction favored by the condensate. In the second block we obtain two interesting new phenomena: a non-universal shear viscosity and a flexoelectric effect. These two effects are due to the anisotropy of our system. 

Anisotropic fluids have been studied in particular in the context of nematic crystals whose hydrodynamic description is given in \cite{LESLIE01011966,Gennes:1974lc}. In this paper, we have initiated the connection of the hydrodynamic description of anisotropic fluids with gauge/gravity duality. The results we obtain in this paper are in agreement with this description, \ie the transport coefficients found here can be related to the ones in \cite{LESLIE01011966}. However since we have not studied the helicity zero modes in much detail, we have not yet described all transport properties. In particular, the thermoelectric effect along the condensate as well as the coefficients $\zeta_x$, $\zeta_y$ and $\lambda$ described in appendix~\ref{sec:General-Remarks-on} are still missing. In the future we plan to study these coefficients in detail. This study may also lead to a covariant hydrodynamic description of anisotropic superfluids.

Furthermore, analytic results close to the phase transition can be found for small values of $\alpha$ \cite{Basu:2011tt}, which determines the ratio of the gravitational to Yang-Mills coupling. On the one hand this analytic approach allows for a detailed study of the transport coefficients close to the phase transition. On the other hand it also permits us to use the fluid/gravity correspondence in order to obtain the complete hydrodynamic description of the system directly from gravity. Similar analyzes for holographic s-wave superfluids can be found in \cite{Herzog:2011ec,Bhattacharya:2011tr,Bhattacharya:2011ee}. We intend to follow this line of thought further.

\section*{Acknowledgements}
We are grateful to A. Buchel, M. Haack, J. Mas, M. Natsuume and G. Policastro for discussions. J.E. is grateful to the KITP Santa Barbara for hospitality during the final stages of this work. This work was supported in part by  {\it The Cluster of Excellence for Fundamental Physics - Origin and Structure of the Universe}.


\begin{appendix}

\section{Holographic Renormalization}
\label{sec:Holographic-Renormalization}
The goal we are pursuing in this section is to find covariant counterterms which can be subtracted from the action~\eqref{eq:hel2action} and \eqref{eq:hel1action} in order to make it finite. We follow the lead of the references \cite{Skenderis:2002wp,Sahoo:2010sp} to perform the holographic renormalization.

\subsection{Asymptotic Behavior}
\label{sec:Asymptotic-Behavior}
In this section we look at the behavior of the fields at the horizon and at the boundary. We want to calculate real-time retarded Green's functions \cite{Son:2002sd,Son:2000xc}, i.e. at the horizon, besides regularity\footnote{Even with all fluctuations switched on, there is no need for a further constraint besides $\phi(r_H)=0$ at the horizon to guarantee regularity.}, we have to fulfill the incoming boundary condition. For this purpose we plug in the ansatz,
  \begin{equation}
  \label{eq:expfluchorpwave}
   F(r)\big|_{r\rightarrow r_H} =\epsilon_h^{\beta} \sum_{i\geq0} \epsilon_h^i F_i^h\,,\quad\text{with}\quad\epsilon_h=\frac{r}{r_h}-1\,,
  \end{equation}
  for the behavior of the fields near the horizon, into the equations of motion~\eqref{eq:hel2eom}, \eqref{eq:hel1eomb1} and \eqref{eq:hel1eomb2}. It turns out that, as expected, we obtain two possibilities for $\beta$, namely
  \begin{equation}
   \beta = \pm i \frac{\omega}{4 \pi T},
  \end{equation}
  with $T$ being the temperature defined in equation ~\eqref{eq:temperature}. As said before, we choose the solution with the ``$-$`` sign which corresponds to the incoming boundary condition. Note that the other solution represents the outgoing boundary condition.
  
  Our ansatz at the boundary is similar to the one used for the background calculation in section~\ref{sec:Holographic-Setup-and}. However, here we have to add a logarithmic term to get a consistent solution (c.f. \cite{Skenderis:2002wp}). Therefore we use
  \begin{equation}
  \label{eq:expflucbdy}
   F(r)\big|_{r\rightarrow r_\text{bdy}} =\sum_{i\ge0}\epsilon_b^i\left(F_i^b+\frac{1}{2}\hat F_i^b\ln\epsilon_b\right)\,,\quad\text{with}\quad \epsilon_b=\left(\frac{r_h}{r}\right)^2\,.
  \end{equation}

Let us now use the above expansions for the helicity one states. In the case of the equations \eqref{eq:hel1eomb1} we have 3 independent expansion coefficients at the boundary (4 free parameters from the 2 second order differential equations minus 1 free parameter due to the constraint). We choose them to be $\left(a_y^3\right)_0^b,\, \left(a_y^3\right)_1^b$ and $\big(\Psi_t\big)_0^b$. At the horizon we already halved the independent parameters by choosing the incoming boundary condition. From the remaining two parameters we can get rid of one by using the constraint equation \eqref{eq:eompwavefluca3ypsit}. Therefore we are left with one free parameter at the horizon, we choose  $\left(a_y^3\right)_0^h$. When solving these equations numerically we set $\left(a_y^3\right)_0^h=1$ and scan through different values of $\omega$.
  
  We can perform similar considerations for the second part (equations \eqref{eq:hel1eomb2}). In this case it is even simpler. We do not have any constraint, just three fields and their corresponding equations of motion. Therefore at the boundary we have six independent parameters, namely $\left(a_y^1\right)_0^b,\, \left(a_y^1\right)_1^b,\, \left(a_y^2\right)_0^b,\, \left(a_y^2\right)_1^b,\, \big(\Psi_x\big)_0^b$ and $\big(\Psi_x\big)_2^b$. At the horizon we have $\big(\Psi_x\big)_0^h,\, \left(a_y^1\right)_0^h,\, \left(a_y^2\right)_0^h$. Note that, as before, we already fixed three free parameters at the horizon by choosing the incoming boundary condition. Again by choosing the values for all fields at the horizon the system is fully determined.
  
  Notice that the same is true for the helicity 2 state. We have again 2 independent components at the boundary, namely $\big(\Xi\big)_0^b$ and $\big(\Xi\big)_2^b$ which are fixed by the incoming boundary condition and $\big(\Xi\big)_0^h$ at the horizon. Therefore as before, the equation is fully determined.
  
  In the following we state the first few non-vanishing coefficients of the expansion at the boundary of the different fields. We need them later on to determine divergences in the on-shell action and to calculate the Green's function. The explicit form of the dependent coefficients are
  \begin{subequations}
   \begin{align}
    \left(\hat a^1_y\right)^b_1 &= \frac{1}{2}\left[\tilde\omega^2 \left(a^1_y\right)^b_0 - 2\ii\tilde\omega \tilde\mu \left(a^2_y\right)^b_0 +\tilde\mu^2 \left(a^1_y\right)^b_0\right]\,,\\
    \left(\hat a^2_y\right)^b_1 &= \frac{1}{2}\left[\tilde\omega^2 \left(a^2_y\right)^b_0 + 2\ii\tilde\omega \tilde\mu \left(a^1_y\right)^b_0 +\tilde\mu^2 \left(a^2_y\right)^b_0\right]\,,\\
    \left(\hat a^3_y\right)^b_1 &= \frac{1}{2}\tilde\omega^2 \left(a^3_y\right)^b_0\,,\\
    \left(\Xi\right)^b_1 &= \frac{1}{4}\tilde\omega^2 \big(\Xi\big)^b_0\,,\qquad \left(\hat \Xi\right)^b_2 = \frac{1}{16}\tilde\omega^4 \big(\Xi\big)^b_0\,,\\
    \big(\Psi_t\big)^b_2 &= -\alpha^2 \tilde\phi^b_2 \left(\tilde a^3_y\right)^b_0\,,\qquad \left(\hat \Psi_t\right)^b_3 = -\frac{\alpha^2 \tilde\omega^2}{3}\tilde\phi^b_2 \left( \tilde a^3_y\right)^b_0\,,\\
    \left(\Psi_x\right)^b_1 &= \frac{1}{4}\tilde\omega^2 \big(\Psi_x\big)^b_0\,,\qquad \left(\hat \Psi_x\right)^b_2 = \frac{1}{16}\tilde\omega^4 \big(\Psi_x\big)^b_0\,.
   \end{align}
  \end{subequations}
  Note that $\mu \equiv \phi^b_0$ and $\phi^b_1$ are the expansion coefficients of $\phi(r)$ at the boundary.
  
  We do not state the expansion at the horizon, since there is no additional information to equation (\ref{eq:expfluchorpwave}), and the explicit form of the non-independent coefficients is very long.

\subsection{Counterterms}
\label{sec:Counterterms}

By plugging the expansions~(\ref{eq:expflucbdy}) into~\eqref{eq:hel2action} and \eqref{eq:hel1action}, resulting in
  \begin{equation}
   \label{eq:actbdryflucpwave}
   \begin{split}
    \frac{\call_{r_b}}{r_h^4}=&{\big(\Xi\big)_0^b}{\big(\Xi\big)_2^b}+\left(-2 f_2^b-\frac{\tilde m_0^b}{2}\right) {\big(\Xi\big)_0^b}^2+{\big(\Psi_x\big)_0^b}{\big(\Psi_x\big)_2^b}+\left(4 f_2^b-\frac{\tilde m_0^b}{2}\right) {\big(\Psi_x\big)_0^b}^2\\
    &-\frac{3}{2}\tilde m_0^b {\big(\Psi_t\big)_0^b}^2+\alpha^2\left[{\left(\tilde a_y^1\right)_0^b}{\left(\tilde a_y^1\right)_1^b}+{\left(\tilde a_y^2\right)_0^b}{\left(\tilde a_y^2\right)_1^b}+{\left(\tilde a_y^3\right)_0^b}{\left(\tilde a_y^3\right)_1^b}\right]\\
    &-\frac{1}{4} \alpha ^2 \tilde \mu ^2 \left[{\left(\tilde a_y^1\right)_0^b}^2 + {\left(\tilde a_y^2\right)_0^b}^2\right]+\alpha^2\left[-\tilde w_2^b{\left(\tilde a_y^1\right)_0^b}{\big(\Psi_x\big)_0^b}+2{\tilde \phi_2^b}{\left(\tilde a_y^3\right)_0^b}{\big(\Psi_t\big)_0^b}\right]\\
    & +\ii\alpha^2\tilde\omega\tilde\mu {\left(\tilde a_y^1\right)_0^b} {\left(\tilde a_y^2\right)_0^b}\\
    &+\left(\frac{1}{8} \epsilon_b \tilde\omega^2+\frac{1}{64} \tilde\omega^4+\frac{1}{32} \tilde\omega^4 \ln\epsilon_b \right)\left[{\big(\Psi\big)_0^b}^2+{\big(\Psi_x\big)_0^b}^2\right]\\
    &-\left(\frac{1}{4} \alpha ^2 \tilde\omega ^2-\frac{1}{4} \alpha^2 \tilde\omega^2\ln\epsilon_b\right)\left[ {\left(\tilde a_y^1\right)_0^b}^2+ {\left(\tilde a_y^2\right)_0^b}^2+{\left(\tilde a_y^3\right)_0^b}^2\right]\\
    &+\frac{1}{4} \alpha^2 \tilde\mu^2\ln\epsilon_b \left[{\left(\tilde a_y^1\right)_0^b}^2+{\left(\tilde a_y^2\right)_0^b}^2\right]\\
    &-\ii\alpha^2\tilde\omega\tilde\mu{\left(\tilde a_y^1\right)_0^b} {\left(\tilde a_y^2\right)_0^b}\ln\epsilon_b \bigg|_{r=r_{\text{bdy}}}.
   \end{split}
  \end{equation}
  Note that $S_\text{on-shell} = \frac{1}{\kappa_5^2}\int\frac{\dd^4 k}{(2\pi)^4}\ \call_{r_b}$ with $r=r_{\text{bdy}} \gg 1$. Moreover, as before, the first field expansion coefficient is always a field of $-k$ and the second of $k$, e.g. ${\big(\Xi\big)_0^b}(-k){\big(\Xi\big)_2^b}(k)$. It is obvious from the last few lines of (\ref{eq:actbdryflucpwave}) that we have to add counter terms to the on-shell action $S_\text{on-shell}$ to take the divergences in $r$ into account.

The terms that have to be considered are the ones in (\ref{eq:actbdryflucpwave}) with explicit $r$ dependence,
\begin{align*}
    &\left(\frac{1}{8} \epsilon_b \tilde\omega^2+\frac{1}{32} \tilde\omega^4 \ln\epsilon_b\right)\left[{\big(\Xi\big)_0^b}^2+{\big(\Psi_x\big)_0^b}^2\right]\, ,\\
    &\frac{1}{4} \alpha^2 \ln\epsilon_b\left\{\tilde\omega^2\left[{\left(\tilde a_y^1\right)_0^b}^2+ {\left(\tilde a_y^2\right)_0^b}^2+{\left(\tilde a_y^3\right)_0^b}^2\right]+\tilde \mu^2 \left[{\left(\tilde a_y^1\right)_0^b}^2+{\left(\tilde a_y^2\right)_0^b}^2\right]\right\}\  \text{ and}\\
    &-\ii\alpha^2\tilde\omega\tilde\mu{\left(\tilde a_y^1\right)_0^b} {\left(\tilde a_y^2\right)_0^b}\ln\epsilon_b
    \end{align*}
    First we need the induced metric $\gamma$ on the $r=r_{\text{bdy}}$ plane. The induced metric is defined by
    \begin{equation}
     \gamma_{\mu\nu} = \frac{\partial x^M}{\partial \tilde{x}^\mu}\frac{\partial x^N}{\partial \tilde{x}^\nu}g_{MN}(r)\bigg|_{r=r_\text{bdy}},
    \end{equation}
    resulting in
    \begin{equation}
     \dd s^2_{r_\text{bdy}} = -N(r_{\text{bdy}})\sigma(r_\text{bdy})^2 \dd t^2 + \frac{r_\text{bdy}^2}{f(r_\text{bdy})^4}\dd x^2 + r_\text{bdy}^2 f(r_\text{bdy})^2(\dd y^2 + \dd z^2).
    \end{equation}
    We do not literally derive the covariant counter terms here in this work. However, by looking at the counter terms of B. Sahoo and H.-U. Yee calculated in \cite{Sahoo:2010sp} we get an idea how they should look like, namely some combinations of $R[\gamma],\, R_{\mu\nu}[\gamma]$ and $F^a_{\mu\nu}$. The first two are the Ricci scalar and Ricci tensor on the induced surface respectively, the latter is the field strength tensor on that surface. Possible covariant combinations of the three terms are $\sqrt{-\gamma}R[\gamma]$, $\sqrt{-\gamma}R^{\mu\nu}[\gamma]R_{\mu\nu}[\gamma]$ and $\sqrt{-\gamma}F^a_{\mu\nu}F^{a\mu\nu}$. Now lets have a look at their expansion for $r \gg 1$.
    \begin{equation}
     \begin{split}
      \sqrt{-\gamma}R[\gamma]\bigg|_{r \gg 1} &= \frac{r^2\omega^2}{2}\left[{\big(\Xi\big)_0^b}^2+{\big(\Psi_x\big)_0^b}^2\right]\, ,\\
      \sqrt{-\gamma}R^{\mu\nu}[\gamma]R_{\mu\nu}[\gamma]\bigg|_{r \gg 1} &= \frac{\omega^4}{2}\left[{\big(\Xi\big)_0^b}^2+{\big(\Psi_x\big)_0^b}^2\right]\ \text{ and}\\
      \sqrt{-\gamma}F^a_{\mu\nu}F^{a\mu\nu}\bigg|_{r \gg 1} &= -2 \bigg\{\omega^2\left[{\left(a_y^1\right)_0^b}^2+ {\left(a_y^2\right)_0^b}^2+{\left(a_y^3\right)_0^b}^2\right]\\&+\mu^2 \left[{\left(a_y^1\right)_0^b}^2+{\left(a_y^2\right)_0^b}^2\right]-4\ii\omega\mu{\left(a_y^1\right)_0^b} {\left(a_y^2\right)_0^b}\bigg\}.
     \end{split}
    \end{equation}
    Therefore by adding the real space action
    \begin{equation}
     S_\text{ct} = - \frac{1}{\k_5^2}\int \dd^4x\ \sqrt{-\gamma}\ \left(\frac{1}{4}R[\gamma]-\frac{1}{16}R^{\mu\nu}[\gamma]R_{\mu\nu}[\gamma]\ln\epsilon_b+\frac{\alpha^2}{8}F^a_{\mu\nu}F^{a\mu\nu}\ln\epsilon_b\right)\bigg|_{r=r_{\text{bdy}}}
    \end{equation}
    to the action $S_\text{on-shell}$~\eqref{eq:hel2action} and \eqref{eq:hel1action}  we get a divergence-free theory (up to the second order in the fluctuations) for $r_{\text{bdy}}\gg 1$, i.e. also the real time Green's functions are divergence free. The renormalized $r_{\text{bdy}}\gg 1$ Lagrangian is then given in~\eqref{eq:hel2actionren} and \eqref{eq:hel1actionren}.

\section{Constructing the Gauge Invariant Fields}
\label{sec:Constructing-the-Gauge}

\subsection{Residual Gauge Transformations}
\label{sec:Residual-Gauge-Transformations}
The transformations we have to look at are diffeomorphisms and $SU(2)$ gauge transformations. On the one hand, we demand that the fields be diffeomorphism invariant, i.e.
    \begin{equation}
     \delta_\Sigma \Phi = \call_\Sigma \Phi = 0.
    \end{equation}
    $\call_\Sigma$ is the Lie derivative along $\Sigma$, i.e.
    \begin{equation}
      \begin{split}
       \call_\Sigma g_{MN} &= \nabla_M \Sigma_N + \nabla_N \Sigma_M = \partial_M \Sigma_N + \partial_N \Sigma_M - 2\Gamma^P_{MN}\Sigma_P,\\
       \call_\Sigma A^a_M &= \Sigma^P \nabla_P A^a_M + A^a_P \nabla_M \Sigma^P = \Sigma^P \partial_P A^a_M + A^a_P \partial_M \Sigma^P,
      \end{split}
    \end{equation}
    with $\Gamma^P_{MN}$ being the Christoffel symbol.\\
    On the other hand they have to be invariant under the $SU(2)$,
    \begin{equation}
     \Phi \to M(\Lambda) \Phi = \Phi,
    \end{equation}
    with $M(\Lambda)$ being the $SU(2)$ transformation matrices, this is equivalent to
    \begin{equation}
     \delta_\Lambda \Phi = 0.
    \end{equation}
    $\Phi$ stands for the physical modes in our system and is composed of the helicity 0 fields $\xi_{tx},\, \xi_{t},\, \xi_{x},\, \xi_{y}, a^a_x$ and $a^a_t$, with $a=1,2,3$. The invariance of $\Phi$ under the above transformations translates into
    \begin{equation}
     \delta \Phi = (\delta_\Sigma + \delta_\Lambda)\Phi = \sum_{a=1}^3 (\tau_a\delta a^a_x + \tau_{3+a}\delta a^a_t) + \tau_7 \delta \xi_{tx} + \tau_8\delta \xi_{t}+\tau_9\delta \xi_{x}+\tau_{10}\delta \xi_{y} = 0, \label{eq:physfieldinv}
    \end{equation}
    with $\tau_i$ being the $r$ dependent coefficients.

\subsubsection{Diffeomorphism Invariance}
\label{sec:Diffeomorphism-Invariance}

    Let us look at the invariance under diffeomorphisms. We begin by defining
    \begin{equation}
     \begin{split}
      \hat{g}_{MN} &= g_{MN} + h_{MN},\\
      \hat{A}^a_M &= A^a_M + a^a_M.
     \end{split}
    \end{equation}
    Furthermore note that $\Sigma_M$ (and later on $\Lambda$) are of the same order as the fluctuations.\\
    Through the gauge choice $h_{Mr}=0$ we can determine the form of $\Sigma_M$ up to some constants, because
    \begin{equation}
     \begin{split}
      &\call_\Sigma \hat{g}_{Mr}=0\\
      \Rightarrow& \partial_M \Sigma_r + \partial_r \Sigma_M - 2\Gamma^P_{Mr}\Sigma_P = 0.
     \end{split}
    \end{equation}
    Note that we just need the Christoffel symbols to zeroth order in fluctuations, i.e. the background Christoffel symbols. They are
    \begin{equation}
     \begin{aligned}
      &\Gamma^r_{rr}=\frac{{c_4}'}{c_4},&&\Gamma^t_{tr}=\frac{{c_1}'}{c_1},&& \Gamma^x_{xr}=\frac{{c_2}'}{c_2},&& \Gamma^y_{yr}=\frac{{c_3}'}{c_3},\\
      &\Gamma^r_{tt}=\frac{c_1{c_1}'}{{c_4}^2},&& \Gamma^r_{xx}=-\frac{c_2{c_2}'}{{c_4}^2}&&\quad \text{ and } && \Gamma^r_{yy}=-\frac{c_3{c_3}'}{{c_4}^2},
     \end{aligned}
    \end{equation}
    with
    \begin{equation}
     \dd s^2 = g_{MN}\dd x^M \dd  y^N = -{c_1(r)}^2 \dd t^2 + {c_2(r)}^2 \dd x^2 + {c_3(r)}^2 (\dd y^2+\dd z^2) + {c_4(r)}^2 \dd r^2.
    \end{equation}
    We get 4 equations (+1 for the $z$ component which is exactly the same as the one for the $y$ component), which read
    \begin{subequations}
     \begin{align}
      &-i\omega \Sigma_r + {\Sigma_t}' - 2\frac{{c_1}'}{c_1}\Sigma_t = 0,\\
      &i k \Sigma_r + {\Sigma_x}' - 2\frac{{c_2}'}{c_2}\Sigma_x = 0,\\
      &{\Sigma_y}' - 2\frac{{c_3}'}{c_3}\Sigma_y = 0,\\
      & 2 {\Sigma_r}' -2\frac{{c_4}'}{c_4}\Sigma_r =0.
     \end{align}
    \end{subequations}
    We work in momentum space, i.e. the ansatz used is
    \begin{equation}
     \Sigma_M(t,x,r) = \int \dd^4 x\ \ee^{ik_\mu x^\mu} \Sigma(\omega,k,r),
    \end{equation}
    with $k^\mu = (\omega,k,0,0)$. The solutions to the equations above are
    \begin{equation}
     \begin{split}
      \Sigma_t(\omega,k,r) &=K_t c_1^2+\ii\omega K_r {c_1}^2 A, \text{ with } A = \int \dd r \frac{c_4}{{c_1}^2},\\
      \Sigma_x(\omega,k,r) &=K_x c_2^2-\ii k K_r {c_2}^2 B, \text{ with } B = \int \dd r \frac{c_4}{{c_2}^2},\\
      \Sigma_y(\omega,k,r) &= K_y {c_3}^2,\\
      \Sigma_r(\omega,k,r) &= K_r {c_4},
     \end{split}
    \end{equation}
    with $K_i$ being constants. Using these solutions in the remaining equations $\delta_\Sigma \xi_i$, we get
    \begin{equation}
     \begin{split}
	\delta_\Sigma \xi_t &=  \frac{2\ii\omega}{{c_1}^2}\Sigma_t,\\
	\delta_\Sigma \xi_x &=  \frac{2\ii k}{{c_2}^2}\Sigma_x,\\
	\delta_\Sigma \xi_y &= 0,\\
	\delta_\Sigma \xi_{tx}&=-\frac{\ii\omega}{{c_2}^2}\Sigma_x +\frac{\ii k}{{c_2}^2}\Sigma_t.
     \end{split}
    \end{equation}
    Here we see already that $\xi_y$ is a physical mode!\\
    Applying the same procedure to the gauge fields, i.e. $\delta_\Sigma a^a_\mu = \call_\Sigma \hat{A}_\mu^a$, it results in
    \begin{equation}
     \begin{split}
      \delta_\Sigma a^1_x &= \frac{w'}{{c_4}^2}\Sigma_r + \frac{ikw}{{c_2}^2}\Sigma_x,\\
      \delta_\Sigma a^3_t &= \frac{\phi'}{{c_4}^2}\Sigma_r + \frac{i\omega \phi}{{c_1}^2}\Sigma_t.\\
     \end{split}
    \end{equation}
    The Lie derivatives of the remaining components vanish.

\subsubsection{$SU(2)$ Gauge Invariance}
\label{sec:$SU(2)$-Gauge-Invariance}
 This transformation only affects the gauge fields, therefore we do not have to care about the metric fluctuations here. A field in the adjoint $SU(2)$ representation transform under the $SU(2)$ as
    \begin{equation}
     \delta_\Lambda a^a_M = \nabla_M \Lambda^a(t,x,r) + \epsilon^{abc}A_M^b \Lambda^c.
    \end{equation}
    Again we constrain the possible $\Lambda^a$ by using the gauge choice $a^a_{r}=0$, i.e.
    \begin{equation}
     \begin{split}
      &0=\delta_\Lambda a^a_r = \nabla_r \Lambda^a(t,x,r) = \partial_r \Lambda^a(t,x,r)\\
      &\Rightarrow \Lambda^a (t,x,r) = \Lambda^a(t,x).
     \end{split}
    \end{equation}
    We choose the ansatz $\Lambda^a(t,x)=\int \dd^4x\ \ee^{ik_\mu x^\mu} \Lambda^a(\omega,k)$ with $k^\mu=(\omega,k,0,0)$ to calculate $\delta_\Lambda a^a_\mu$. Note that by using the definition above of $k^\mu$ the $y$ and $z$ components do not mix with the rest and we can forget about them. We end up with
    \begin{equation}
    \begin{aligned}
     &\delta_\Lambda a^1_x = \ii k\Lambda^1, && \delta_\Lambda a^1_t = -\ii\omega\Lambda^1-\phi\Lambda^2,\\
     &\delta_\Lambda a^2_x = \ii k\Lambda^2 - w\Lambda^3, && \delta_\Lambda a^2_t = -\ii \omega\Lambda^2+\phi\Lambda^1,\\
     &\delta_\Lambda a^3_x = \ii k\Lambda^3 + w\Lambda^2, && \delta_\Lambda a^3_t = -\ii \omega\Lambda^3.\\
    \end{aligned}
    \end{equation}

\subsection{Physical Fields}
\label{sec:Physical-Fields}
Plugging everything into equation (\ref{eq:physfieldinv}) results in 6 equations, due to the fact that $\Sigma_t,\,\Sigma_x,\,\Sigma_r,\,\Lambda^1,\,\Lambda^2$ and $\Lambda^3$ are independent. The equations are
    \begin{equation}
     \begin{split}
      0=&\ii k\tau_1 -\ii \omega \tau_4 + \phi \tau_5,\\
      0=&\ii k\tau_2 +w \tau_3 -\phi \tau_4 -\ii\omega \tau_5,\\
      0=&-w \tau_2 +\ii k\tau_3-\ii\omega\tau_6,\\
      0=&-\ii\omega \tau_7 + 2\ii k\tau_9+\ii kw\tau_1,\\
      0=&2\ii\omega\tau_8 + \ii k\frac{{c_1}^2}{{c_2}^2}\tau_7+\ii\omega\phi \tau_6,\\
      0=&w'\tau_1 + \phi' \tau_6.
     \end{split}
    \end{equation}
The four physical fields, $\Phi_i = \sum_{i=1}^{10}\tau_i\cdot (\text{Helicity 0 fields})$, we get by solving above equations, are
    \begin{equation}
     \begin{split}
      \Phi_1 =& \xi_y,\\
      \Phi_2 =&a^1_t+\frac{i \omega}{\phi}a^2_t+\frac{i k \left(\omega ^2-\phi^2\right)}{\left(k^2-w^2\right) \phi}a^2_x+\frac{w \left(\omega ^2-\phi^2\right)}{\left(k^2-w^2\right) \phi}a^3_x,\\
      \Phi_3 =&\xi_x-\frac{k^2 {c_1}^2}{\omega ^2 {c_2}^2}\xi_t+\frac{2 k}{\omega}\xi_{tx},\\
      \Phi_4 =&a^1_x+\frac{k}{\omega}a^1_t-\frac{1}{2} w \xi_x-\frac{w'}{\phi'}a^3_t+\frac{\phi w'}{2 \phi'}\xi_t-\frac{k \left(\omega ^2 w'+w \phi \phi'\right)}{\omega  \left(k^2-w^2\right) \phi'}a^3_x\\&-\ii\frac{\omega^2ww'+k^2\phi\phi'}{\omega\phi'(k^2-w^2)}a^2_x .
     \end{split}\label{eq:physicalfields}
    \end{equation}

\section{Numerical Evaluation of Green's Functions}
\label{sec:Numerical-Evalutation-of}
Here we review the algorithm to evaluate Green's function when there is operator mixing \cite{Kaminski:2009dh}. For concreteness we present the algorithm for the second block of the helicity one fields where we have the mixing of the fluctuations $a^1_y,a^2_y,\Psi_x$. Let us first go back to the action~\eqref{eq:hel1action} and have a look at the components we need,
   \begin{equation}
    \begin{split}
      S_\text{on-shell} &\supset \frac{1}{\kappa_5^2}\int \frac{\dd^4 k}{(2\pi)^4}\ \bigg\{
      -\frac{1}{4} r^3 f^6 N \sigma {\Psi_x} {\Psi_x}'-\frac{r \alpha ^2 N \sigma}{2 f^2}{a^1_y}{a^1_y}'-\frac{r \alpha ^2 N \sigma}{2 f^2}{a^2_y}{a^2_y}'\\
      &+\left(\frac{3}{2} r^3 f^6 \sqrt{N} \sigma-r^2 f^6 N \sigma-r^3 f^5 N \sigma f'-\frac{1}{4} r^3 f^6 \sigma N'-\frac{1}{2} r^3 f^6 N \sigma'\right){\Psi_x}^2\\
      &+\frac{r \alpha ^2 f^4 N \sigma w'}{2} {a^1_y}{\Psi_x}\bigg\}\bigg|_{r=r_{\text{bdy}}}=\\
      &=\frac{1}{\kappa_5^2}\int \frac{\dd^4 k}{(2\pi)^4}\ \left(\Lambda^\text{T}(-k,r) A(k,r) \partial_r\Lambda(k,r) + \Lambda^\text{T}(-k,r) B(k,r) \Lambda(k,r)\right)\bigg|_{r=r_\text{bdy}}.
    \end{split}
   \end{equation}
   $\Lambda^\text{T}(k,r)=\left(a^1_y(k,r),a^2_y(k,r),\Psi_x(k,r)\right)$, $A(k,r)$ and $B(k,r)$ are the matrices containing the coefficients of the field's bilinear forms. Their explicit form for $r=r_{\text{bdy}}\gg1$ is
   \begin{align}
    A(k,r_\text{bdy}) = \left(\begin{array}{ccc}
     -\frac{\alpha^2r_{\text{bdy}}^3}{2} & 0 & 0 \\
      0 & -\frac{\alpha^2r_{\text{bdy}}^3}{2} &0 \\
      0 & 0 & -\frac{r_{\text{bdy}}^5}{4}
    \end{array}\right),\, &
    B(k,r_{\text{bdy}}) = \left(\begin{array}{ccc}
     0 & 0 & -\alpha^2 w_1^br_h^2 \\
      0 & 0 &0 \\
      0 & 0 & 4f^b_2r_h^4
    \end{array}\right).
   \end{align}
   Note that all components of $B(k,r_{\text{bdy}})$ are contact terms.
   
   The next step is to define the boundary condition at the horizon for the vector $\Lambda(k,r_h)$. This includes the incoming boundary condition we already introduced before (c.f. eq. (\ref{eq:expfluchorpwave})). We define
   \begin{equation}
    \Lambda_{(a)}^I(k,r\rightarrow r_H) \simeq \epsilon_h^{-\frac{i\omega}{4\pi T}}\left(e^I_{(a)} + \calo\left(\epsilon_h\right)\right),
   \end{equation}
   the index $I$ refers to the three fields $a^1_y,\, a^2_y$ and $\Psi_x$. Due to the fact that we have 3 coupled second order differential equations we need 6 independent boundary conditions. We already halved them by demanding incoming boundary conditions. We determine the remaining 3 conditions by choosing 3 linear independent vectors $e_{(1)},\, e_{(2)}$ and $e_{(3)}$, with
   \begin{equation}
    e_{(1)}^\text{T} = (1,0,0)\,,\quad
    e_{(2)}^\text{T} = (0,1,0)\,,\quad
    e_{(3)}^\text{T} = (0,0,1)\,.
   \end{equation}
   Note that alternate choices are possible, but we get the best numerical result (with least noise) using these values. We are now able to generate three linearly independent solutions for the equations of motion by solving them numerically using the three linearly independent boundary conditions. We put them together in a 3 by 3 matrix $H(k,r)$ defined by
   \begin{equation}
    H^I_{~a} (k,z) = \Lambda^I_{(a)}(k,r),
   \end{equation}
   with $I=1,2,3$ and $a=1,2,3$. Now we can define a further matrix 
   \begin{equation}
\label{eq:matrixF}
   \calh(k,r)=H(k,r) H(k,r_{\text{bdy}})^{-1}\,.
\end{equation}
  This matrix is basically the fields divided by the field's value at the boundary, i.e. if we had decoupled differential equations we would get $\calh=\diag\left(\Lambda^1(r)/\Lambda^1(r_{\text{bdy}}),\Lambda^2(r)/\Lambda^2(r_{\text{bdy}}),...\right)$.\\
   Next we take the derivative with respect to $r$ of this matrix, leading us to $\calh'(k,r)=H'(k,r)\cdot H(k,r_{\text{bdy}})^{-1}$. Again looking at a possible decoupled case we would obtain $\calh'(r)=\diag\left({\Lambda^1}'(r)/\Lambda^1(r_{\text{bdy}}),{\Lambda^2}'(r)/\Lambda^2(r_{\text{bdy}}),\ldots\right)$ which took the ratio between the derivative of the field and the field into account. This is the know result \cite{Son:2002sd,Herzog:2002pc}.  What is missing for the Green's function is the $r$ dependent prefactor, which is exactly the content of $A(k,r)$. Therefore we finally get
   \begin{equation}
    G^R(\omega) = 2 \lim_{r_B\to\infty} A(\omega,r_{\text{bdy}})\calh'(\omega,r_{\text{bdy}}) + \text{contact- and counter terms}\,,
   \end{equation}
   which allows us to calculate the retarded Green's function in systems where the operators may mix.
\section{General Remarks on Viscosity in Anisotropic Fluids}
\label{sec:General-Remarks-on}
In general, viscosity refers to the dissipation of energy due to any
internal motion \cite{Landau:fluid}. For an internal motion which
describes a general translation or a general rotation, the dissipation
is zero. Thus the dissipation depends  on the gradient of the
velocities $u^\mu$ only  in the combination $u_{\mu\nu}=\frac{1}{2}\left(\nabla_\mu u_\nu+\nabla_\nu u_\mu\right)$, and we may define a general dissipation function $\Xi=\frac{1}{2}\eta^{\mu\nu\lambda\rho}u_{\mu\nu}u_{\lambda\rho}$, where $\eta^{\mu\nu\lambda\rho}$ defines the viscosity tensor \cite{Landau:1959te}. Its symmetries are given by
\begin{equation}
\label{eq:symeta}
\eta^{\mu\nu\lambda\rho}=\eta^{\nu\mu\lambda\rho}=\eta^{\mu\nu\rho\lambda}=\eta^{\lambda\rho\mu\nu}\,.
\end{equation}
The part of the stress tensor which is dissipative 
due to viscosity is defined by
\begin{equation}
\label{eq:stress}
\Pi^{\mu\nu}=-\frac{\del \Xi}{\del u_{\mu\nu}}=-\eta^{\mu\nu\lambda\rho}u_{\lambda\rho}\,.
\end{equation}
We consider a fluid in the rest frame of the normal fluid $u^t=1$. To
satisfy the condition of the Landau frame $u_\mu
\Pi^{\mu\nu}=0$, the stress energy tensor and thus the
viscosity has non-zero components only in the spatial directions $i,j=\{x,y,z\}$. In general only 21 independent components of $\eta_{ijkl}$ appear in the expressions above.

For an isotropic fluid, there are only two independent components which are usually parametrized by the shear viscosity $\eta$ and the bulk viscosity $\zeta$. The dissipative part of the stress tensor becomes $\Pi^{ij}=-2\eta(u^{ij}-\frac{1}{3}\delta^{ij}u_{l}^l)-\zeta u_{l}^l\delta^{ij}$ which is the well-known result.

In a transversely isotropic fluid, there are five independent components of the tensor $\eta^{ijkl}$. For concreteness we choose the symmetry axis to be along the $x$-axis. The non-zero components are given by
\begin{equation}
\label{eq:etatensorpwave}
\begin{aligned}
&\eta^{xxxx}=\zeta_x-2\lambda\,,
&&\eta^{yyyy}=\eta^{zzzz}=\zeta_y-\frac{\lambda}{2}+\eta_{yz}\,,\\
&\eta^{xxyy}=\eta^{xxzz}=\lambda\,,
&&\eta^{yyzz}=\zeta_y-\frac{\lambda}{2}-\eta_{yz}\,,\\
&\eta^{yzyz}=\eta_{yz}\,,
&&\eta^{xyxy}=\eta^{xzxz}=\eta_{xy}\,.
\end{aligned}
\end{equation}
The non-zero off-diagonal components of the stress tensor are given by
\begin{equation}
\label{eq:off-diagonalstressanisotroptic}
\begin{split}
&\Pi^{xy}=-2\eta_{xy} u_{xy}\,,\quad \Pi^{xz}=-2\eta_{xy} u_{xz}\,,\\
&\Pi^{yz}=-2\eta_{yz} u_{yz}\,,
\end{split}
\end{equation}

So far, we only considered the contribution to the stress tensor due
to the dissipation via viscosity and found the terms in the
constitutive equation which contain the velocity of the normal fluid
$u_\mu$. In general, also terms depending on the derivative of
Nambu-Goldstone boson fields $v_\mu=\del_\mu \varphi$, on the
superfluid velocity and on the velocity of the director may contribute
to the dissipative part of the stress tensor. Here the director is given by the vector pointing in the preferred direction. However these terms do not contribute to the off-diagonal components of the energy-momentum tensor for the following reasons: (1) a shear viscosity due to the superfluid velocity leads to a non-positive divergence of the entropy current \cite{Landau:fluid,Pujol:2002na} and (2) no rank two tensor can be formed out of degrees of freedom of the director if the gradients of the director vanish \cite{LESLIE01011966}. In our case the second argument is fulfilled since the condensate is homogeneous and the fluctuations depend only on time. These degrees of freedom will generate additional transport coefficients, but they do not change the shear viscosities. Thus we can write Kubo formulae which determine the shear viscosities in terms of the stress energy correlation functions.

\end{appendix}

\providecommand{\href}[2]{#2}\begingroup\raggedright\endgroup


\begin{thebibliography}{10}

\bibitem{Erdmenger:2008rm}
J.~Erdmenger, M.~Haack, M.~Kaminski, and A.~Yarom, \emph{ {Fluid dynamics of
  R-charged black holes}}, JHEP {\bf 01} (2009) 055,
\href{http://www.slac.stanford.edu/spires/find/hep/www?texkey=Erdmenger:2008rm}{arXiv:0809.2488}.

\bibitem{Banerjee:2008th}
N.~Banerjee, J.~Bhattacharya, S.~Bhattacharyya, S.~Dutta, R.~Loganayagam, {\em
  et al.}, \emph{ {Hydrodynamics from charged black branes}}, JHEP {\bf 1101}
  (2011) 094,
  \href{http://www.slac.stanford.edu/spires/find/hep/www?texkey=Banerjee:2008th}{arXiv:0809.2596}.

\bibitem{Son:2009tf}
D.~T. Son and P.~Surowka, \emph{ {Hydrodynamics with Triangle Anomalies}},
  Phys.Rev.Lett. {\bf 103} (2009) 191601,
  \href{http://www.slac.stanford.edu/spires/find/hep/www?texkey=Son:2009tf}{arXiv:0906.5044}.

\bibitem{Lifschytz:2009si}
G.~Lifschytz and M.~Lippert, \emph{ {Anomalous conductivity in holographic
  QCD}}, Phys. Rev. {\bf D80} (2009) 066005,
\href{http://www.slac.stanford.edu/spires/find/hep/www?texkey=Lifschytz:2009si}{arXiv:0904.4772}.

\bibitem{Gynther:2010ed}
A.~Gynther, K.~Landsteiner, F.~Pena-Benitez, and A.~Rebhan, \emph{ {Holographic
  Anomalous Conductivities and the Chiral Magnetic Effect}}, JHEP {\bf 02}
  (2011) 110,
\href{http://www.slac.stanford.edu/spires/find/hep/www?texkey=Gynther:2010ed}{arXiv:1005.2587}.

\bibitem{Kalaydzhyan:2011vx}
T.~Kalaydzhyan and I.~Kirsch, \emph{ {Fluid/gravity model for the chiral
  magnetic effect}}, Phys. Rev. Lett. {\bf 106} (2011) 211601,
\href{http://www.slac.stanford.edu/spires/find/hep/www?texkey=Kalaydzhyan:2011vx}{arXiv:1102.4334}.

\bibitem{Hoyos:2011us}
C.~Hoyos, T.~Nishioka, and A.~O'Bannon, \emph{ {A Chiral Magnetic Effect from
  AdS/CFT with Flavor}},
\href{http://www.slac.stanford.edu/spires/find/hep/www?texkey=Hoyos:2011us}{arXiv:1106.4030}.

\bibitem{Mateos:2011tv}
D.~Mateos and D.~Trancanelli, \emph{ {Thermodynamics and Instabilities of a
  Strongly Coupled Anisotropic Plasma}}, JHEP {\bf 07} (2011) 054,
\href{http://www.slac.stanford.edu/spires/find/hep/www?texkey=Mateos:2011tv}{arXiv:1106.1637}.

\bibitem{Rebhan:2011ke}
A.~Rebhan and D.~Steineder, \emph{ {Electromagnetic signatures of a strongly
  coupled anisotropic plasma}}, JHEP {\bf 1108} (2011) 153,
  \href{http://www.slac.stanford.edu/spires/find/hep/www?texkey=Rebhan:2011ke}{arXiv:1106.3539}.

\bibitem{Herzog:2011ec}
C.~P. Herzog, N.~Lisker, P.~Surowka, and A.~Yarom, \emph{ {Transport in
  holographic superfluids}}, JHEP {\bf 1108} (2011) 052,
  \href{http://www.slac.stanford.edu/spires/find/hep/www?texkey=Herzog:2011ec}{arXiv:1101.3330}.

\bibitem{Bhattacharya:2011ee}
J.~Bhattacharya, S.~Bhattacharyya, and S.~Minwalla, \emph{ {Dissipative
  Superfluid dynamics from gravity}}, JHEP {\bf 1104} (2011) 125,
  \href{http://www.slac.stanford.edu/spires/find/hep/www?texkey=Bhattacharya:2011ee}{arXiv:1101.3332}.

\bibitem{Bhattacharya:2011tr}
J.~Bhattacharya, S.~Bhattacharyya, S.~Minwalla, and A.~Yarom, \emph{ {A Theory
  of first order dissipative superfluid dynamics}},
  \href{http://www.slac.stanford.edu/spires/find/hep/www?texkey=Bhattacharya:2011tr}{arXiv:1105.3733}.

\bibitem{Landau:1959te}
L.~D. Landau and E.~M. Lifshitz, {\em { Course of Theoretical Physics, Volume
  7, Theory of Elasticity}}.
\newblock Pergamon Press (1959) 134p.

\bibitem{Gennes:1974lc}
P.~de~Gennes, {\em {The Physics of Liquid Crystals}}.
\newblock Oxford University Press, 1974.

\bibitem{Gubser:2008px}
S.~S. Gubser, \emph{ {Breaking an Abelian gauge symmetry near a black hole
  horizon}}, Phys. Rev. {\bf D78} (2008) 065034,
\href{http://www.slac.stanford.edu/spires/find/hep/www?texkey=Gubser:2008px}{arXiv:0801.2977}.

\bibitem{Hartnoll:2008vx}
S.~A. Hartnoll, C.~P. Herzog, and G.~T. Horowitz, \emph{ {Building a
  Holographic Superconductor}}, Phys. Rev. Lett. {\bf 101} (2008) 031601,
\href{http://www.slac.stanford.edu/spires/find/hep/www?texkey=Hartnoll:2008vx}{arXiv:0803.3295}.

\bibitem{Hartnoll:2008kx}
S.~A. Hartnoll, C.~P. Herzog, and G.~T. Horowitz, \emph{ {Holographic
  Superconductors}}, JHEP {\bf 12} (2008) 015,
\href{http://www.slac.stanford.edu/spires/find/hep/www?texkey=Hartnoll:2008kx}{arXiv:0810.1563}.

\bibitem{Gubser:2008wv}
S.~S. Gubser and S.~S. Pufu, \emph{ {The gravity dual of a p-wave
  superconductor}}, JHEP {\bf 11} (2008) 033,
\href{http://www.slac.stanford.edu/spires/find/hep/www?texkey=Gubser:2008wv}{arXiv:0805.2960}.

\bibitem{Ammon:2008fc}
M.~Ammon, J.~Erdmenger, M.~Kaminski, and P.~Kerner, \emph{ {Superconductivity
  from gauge/gravity duality with flavor}}, Phys. Lett. {\bf B680} (2009)
  516--520,
\href{http://www.slac.stanford.edu/spires/find/hep/www?texkey=Ammon:2008fc}{arXiv:0810.2316}.

\bibitem{Basu:2008bh}
P.~Basu, J.~He, A.~Mukherjee, and H.-H. Shieh, \emph{ {Superconductivity from
  D3/D7: Holographic Pion Superfluid}}, JHEP {\bf 11} (2009) 070,
\href{http://www.slac.stanford.edu/spires/find/hep/www?texkey=Basu:2008bh}{arXiv:0810.3970}.

\bibitem{Ammon:2009fe}
M.~Ammon, J.~Erdmenger, M.~Kaminski, and P.~Kerner, \emph{ {Flavor
  Superconductivity from Gauge/Gravity Duality}}, JHEP {\bf 10} (2009) 067,
\href{http://www.slac.stanford.edu/spires/find/hep/www?texkey=Ammon:2009fe}{arXiv:0903.1864}.

\bibitem{Ammon:2009xh}
M.~Ammon, J.~Erdmenger, V.~Grass, P.~Kerner, and A.~O'Bannon, \emph{ {On
  Holographic p-wave Superfluids with Back-reaction}}, Phys. Lett. {\bf B686}
  (2010) 192--198,
\href{http://www.slac.stanford.edu/spires/find/hep/www?texkey=Ammon:2009xh}{arXiv:0912.3515}.

\bibitem{Kovtun:2004de}
P.~Kovtun, D.~T. Son, and A.~O. Starinets, \emph{ {Viscosity in strongly
  interacting quantum field theories from black hole physics}}, Phys. Rev.
  Lett. {\bf 94} (2005) 111601,
\href{http://www.slac.stanford.edu/spires/find/hep/www?texkey=Kovtun:2004de}{arXiv:hep-th/0405231}.

\bibitem{Buchel:2003tz}
A.~Buchel and J.~T. Liu, \emph{ {Universality of the shear viscosity in
  supergravity}}, Phys. Rev. Lett. {\bf 93} (2004) 090602,
\href{http://www.slac.stanford.edu/spires/find/hep/www?texkey=Buchel:2003tz}{arXiv:hep-th/0311175}.

\bibitem{Iqbal:2008by}
N.~Iqbal and H.~Liu, \emph{ {Universality of the hydrodynamic limit in AdS/CFT
  and the membrane paradigm}}, Phys. Rev. {\bf D79} (2009) 025023,
\href{http://www.slac.stanford.edu/spires/find/hep/www?texkey=Iqbal:2008by}{arXiv:0809.3808}.

\bibitem{Buchel:2008vz}
A.~Buchel, R.~C. Myers, and A.~Sinha, \emph{ {Beyond eta/s = 1/4pi}}, JHEP {\bf
  03} (2009) 084,
\href{http://www.slac.stanford.edu/spires/find/hep/www?texkey=Buchel:2008vz}{arXiv:0812.2521}.

\bibitem{Cremonini:2011iq}
S.~Cremonini, \emph{ {The Shear Viscosity to Entropy Ratio: A Status Report}},
  Mod. Phys. Lett. {\bf B25} (2011) 1867--1888,
\href{http://www.slac.stanford.edu/spires/find/hep/www?texkey=Cremonini:2011iq}{arXiv:1108.0677}.

\bibitem{Erdmenger:2010xm}
J.~Erdmenger, P.~Kerner, and H.~Zeller, \emph{ {Non-universal shear viscosity
  from Einstein gravity}}, Phys.Lett. {\bf B699} (2011) 301--304,
  \href{http://www.slac.stanford.edu/spires/find/hep/www?texkey=Erdmenger:2010xm}{arXiv:1011.5912}.

\bibitem{Natsuume:2010ky}
M.~Natsuume and M.~Ohta, \emph{ {The Shear viscosity of holographic
  superfluids}}, Prog.Theor.Phys. {\bf 124} (2010) 931--951,
  \href{http://www.slac.stanford.edu/spires/find/hep/www?texkey=Natsuume:2010ky}{arXiv:1008.4142}.

\bibitem{Hartnoll:2009sz}
S.~A. Hartnoll, \emph{ {Lectures on holographic methods for condensed matter
  physics}}, Class. Quant. Grav. {\bf 26} (2009) 224002,
\href{http://www.slac.stanford.edu/spires/find/hep/www?texkey=Hartnoll:2009sz}{arXiv:0903.3246}.

\bibitem{Kobayashi:2006sb}
S.~Kobayashi, D.~Mateos, S.~Matsuura, R.~C. Myers, and R.~M. Thomson, \emph{
  {Holographic phase transitions at finite baryon density}}, JHEP {\bf 02}
  (2007) 016,
\href{http://www.slac.stanford.edu/spires/find/hep/www?texkey=Kobayashi:2006sb}{arXiv:hep-th/0611099}.

\bibitem{Balasubramanian:1999re}
V.~Balasubramanian and P.~Kraus, \emph{ {A stress tensor for anti-de Sitter
  gravity}}, Commun. Math. Phys. {\bf 208} (1999) 413--428,
\href{http://www.slac.stanford.edu/spires/find/hep/www?texkey=Balasubramanian:1999re}{arXiv:hep-th/9902121}.

\bibitem{deHaro:2000xn}
S.~de~Haro, S.~N. Solodukhin, and K.~Skenderis, \emph{ Holographic
  reconstruction of spacetime and renormalization in the AdS/CFT
  correspondence}, Commun. Math. Phys. {\bf 217} (2001) 595--622,
\href{http://www.slac.stanford.edu/spires/find/hep/www?texkey=deHaro:2000xn}{arXiv:hep-th/0002230}.

\bibitem{Basu:2009vv}
P.~Basu, J.~He, A.~Mukherjee, and H.-H. Shieh, \emph{ {Hard-gapped Holographic
  Superconductors}}, Phys. Lett. {\bf B689} (2010) 45--50,
\href{http://www.slac.stanford.edu/spires/find/hep/www?texkey=Basu:2009vv}{arXiv:0911.4999}.

\bibitem{Gubser:2010dm}
S.~S. Gubser, F.~D. Rocha, and A.~Yarom, \emph{ {Fermion correlators in
  non-abelian holographic superconductors}}, JHEP {\bf 1011} (2010) 085,
  \href{http://www.slac.stanford.edu/spires/find/hep/www?texkey=Gubser:2010dm}{arXiv:1002.4416}.

\bibitem{Son:2002sd}
D.~T. Son and A.~O. Starinets, \emph{ {Minkowski-space correlators in AdS/CFT
  correspondence: Recipe and applications}}, JHEP {\bf 09} (2002) 042,
\href{http://www.slac.stanford.edu/spires/find/hep/www?texkey=Son:2002sd}{arXiv:hep-th/0205051}.

\bibitem{Herzog:2009xv}
C.~P. Herzog, \emph{ {Lectures on Holographic Superfluidity and
  Superconductivity}}, J.Phys.A {\bf A42} (2009) 343001,
  \href{http://www.slac.stanford.edu/spires/find/hep/www?texkey=Herzog:2009xv}{arXiv:0904.1975}.

\bibitem{Myers:2007we}
R.~C. Myers, A.~O. Starinets, and R.~M. Thomson, \emph{ {Holographic spectral
  functions and diffusion constants for fundamental matter}}, JHEP {\bf 11}
  (2007) 091,
\href{http://www.slac.stanford.edu/spires/find/hep/www?texkey=Myers:2007we}{arXiv:0706.0162}.

\bibitem{Buchel:2009tt}
A.~Buchel and R.~C. Myers, \emph{ {Causality of Holographic Hydrodynamics}},
  JHEP {\bf 08} (2009) 016,
\href{http://www.slac.stanford.edu/spires/find/hep/www?texkey=Buchel:2009tt}{arXiv:0906.2922}.

\bibitem{Basu:2011tt}
P.~Basu and J.-H. Oh, \emph{ {Analytic Approaches to An-Isotropic Holographic
  Superfluids}},
  \href{http://www.slac.stanford.edu/spires/find/hep/www?texkey=Basu:2011tt}{arXiv:1109.4592}.

\bibitem{Erdmenger:2008yj}
J.~Erdmenger, M.~Kaminski, P.~Kerner, and F.~Rust, \emph{ {Finite baryon and
  isospin chemical potential in AdS/CFT with flavor}}, JHEP {\bf 11} (2008)
  031,
\href{http://www.slac.stanford.edu/spires/find/hep/www?texkey=Erdmenger:2008yj}{arXiv:0807.2663}.

\bibitem{Barisch:2011TA}
S.~Barisch, M.~Haack, S.~Nampuri, and G.~Policastro, \emph{ to appear},.

\bibitem{LESLIE01011966}
F.~M. Leslie, \emph{ {Some Constitutive Equations For Anisotropic Fluids}}, The
  Quarterly Journal of Mechanics and Applied Mathematics {\bf 19} (1966),
  no.~3, 357--370.

\bibitem{Skenderis:2002wp}
K.~Skenderis, \emph{ {Lecture notes on holographic renormalization}}, Class.
  Quant. Grav. {\bf 19} (2002) 5849--5876,
\href{http://www.slac.stanford.edu/spires/find/hep/www?texkey=Skenderis:2002wp}{arXiv:hep-th/0209067}.

\bibitem{Sahoo:2010sp}
B.~Sahoo and H.-U. Yee, \emph{ {Electrified plasma in AdS/CFT correspondence}},
  JHEP {\bf 11} (2010) 095,
\href{http://www.slac.stanford.edu/spires/find/hep/www?texkey=Sahoo:2010sp}{arXiv:1004.3541}.

\bibitem{Son:2000xc}
D.~T. Son and M.~A. Stephanov, \emph{ {QCD at finite isospin density}}, Phys.
  Rev. Lett. {\bf 86} (2001) 592--595,
\href{http://www.slac.stanford.edu/spires/find/hep/www?texkey=Son:2000xc}{arXiv:hep-ph/0005225}.

\bibitem{Kaminski:2009dh}
M.~Kaminski, K.~Landsteiner, J.~Mas, J.~P. Shock, and J.~Tarrio, \emph{
  {Holographic Operator Mixing and Quasinormal Modes on the Brane}}, JHEP {\bf
  02} (2010) 021,
\href{http://www.slac.stanford.edu/spires/find/hep/www?texkey=Kaminski:2009dh}{arXiv:0911.3610}.

\bibitem{Herzog:2002pc}
C.~P. Herzog and D.~T. Son, \emph{ {Schwinger-Keldysh propagators from AdS/CFT
  correspondence}}, JHEP {\bf 03} (2003) 046,
\href{http://www.slac.stanford.edu/spires/find/hep/www?texkey=Herzog:2002pc}{arXiv:hep-th/0212072}.

\bibitem{Landau:fluid}
L.~D. Landau and E.~M. Lifshitz, {\em { Course of Theoretical Physics, Volume
  6, Fluid Mechanics}}.
\newblock Pergamon Press (1959) 134p.

\bibitem{Pujol:2002na}
C.~Pujol and D.~Davesne, \emph{ {Relativistic dissipative hydrodynamics with
  spontaneous symmetry breaking}}, Phys. Rev. {\bf C67} (2003) 014901,
\href{http://www.slac.stanford.edu/spires/find/hep/www?texkey=Pujol:2002na}{arXiv:hep-ph/0204355}.

\end{thebibliography}

\end{document}